\documentclass[twocolumn,showpacs,preprintnumbers,amsmath,amssymb,prb]{revtex4}
\usepackage{graphicx}
\usepackage{dcolumn}
\usepackage{bm}
\usepackage{amssymb}
\usepackage{amsmath}
\usepackage{epsf}
\usepackage{amssymb}
\usepackage{amsmath}
\usepackage{epsf}
\usepackage{graphics}
\usepackage{verbatim}
\def\be{\begin{equation}}
\def\ee{\end{equation}}

\def\bee{\begin{eqnarray}}
\def\eee{\end{eqnarray}}

\usepackage[usenames]{color}

\begin{document}

\title{
High-temperature Aharonov-Bohm-Casher interferometer }

\author{P. M. Shmakov$^{1}$}
\author{A. P. Dmitriev$^{1,2}$}
\author{V. Yu. Kachorovskii$^{1}$}
\affiliation{$^{1}$A.F.Ioffe Physico-Technical Institute,
194021 St.~Petersburg, Russia
\\
$^{2}$Institut f\"ur Nanotechnologie,  Karlsruhe Institute of Technology,
76021 Karlsruhe, Germany
}

\date{\today}

\begin{abstract}
{We study theoretically the combined effect of the spin-orbit
and Zeeman interactions on the tunneling
electron transport through a single-channel quantum ring threaded by magnetic flux. We focus on the high temperature case
(temperature is much higher than the level spacing in the ring) and demonstrate that   spin-interference effects are not suppressed by thermal averaging. In the absence of the  Zeeman coupling the high-temperature tunneling conductance  of the ring exhibits  two types of oscillations: Aharonov-Bohm oscillations with magnetic flux and Aharonov-Casher oscillations with the strength of the spin-orbit interaction.  For weak tunneling coupling  both oscillations have the form of sharp periodic antiresonances.
In the vicinity of the antiresonances the tunneling electrons acquire spin polarization, so that the ring serves as a spin polarizer.
We also demonstrate that      the Zeeman coupling  leads to appearance of two additional  peaks    both in the tunneling conductance and in the spin polarization.  }
\end{abstract}
\maketitle

\section{Introduction}

Quantum interferometers based on low-dimensional electronic nanosystems proved to be very powerful tools
in studying coherent  mesoscopic phenomena  \cite{so1,so2,so3,so5}.
 The simplest example of such an interferometer   is a {\it single-channel  ballistic} quantum ring tunnel-coupled
 to leads  and threaded by the magnetic flux $\Phi$ (see Fig.~1).
  The
  dependence of the
   conductance of this setup  $G$   on  $\Phi$ encodes important information about the phase coherence
   of the tunneling   electrons.
  In particular, the interference of electron waves, propagating in the ring
clockwise and counterclockwise,
results in
the Aharonov-Bohm (AB) oscillations  \cite{bohm,aronov}     of
$G$  with $\Phi$.
The oscillation  period is given by the flux quantum $\Phi_0 = hc/e.$
\begin{figure}[ht!]
 \vspace{3.5mm}
 \leavevmode \epsfxsize=5.0cm \centering{\epsfbox{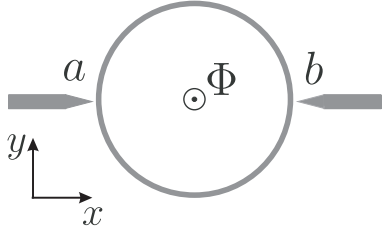}}
\caption{The ring threaded by magnetic flux $\Phi.$
\vspace{3.5mm}
}
\label{fig1}
\end{figure}
The
shape and amplitude of  the AB oscillations
depend essentially on the strength of the tunneling coupling
and on the relation between the  temperature $T$ and the level spacing in the ring $\Delta .$
  For  $T \ll \Delta$ and weak tunneling coupling  there are narrow resonant peaks in the dependence $G(\Phi)$.\cite{butt}
  The positions of the peaks depend on the electron Fermi energy \cite{butt} and the strength of the electron-electron interaction. \cite{kin}
       Remarkably,  the interference effects are not entirely suppressed  by thermal averaging, and the resonant behavior of $G(\Phi)$ survives
for the case $T\gg \Delta.$
Specifically, the high-temperature conductance of the  noninteracting  ring  with  weak tunnel coupling to the contacts exhibits sharp antiresonances at $\phi = 1/2 +N,$  where $\phi=\Phi / \Phi_0$ is the dimensionless flux and $N$ is an arbitrary integer number (see Fig.2). \cite{jagla,dmitriev}  The electron-electron interaction leads to  appearance of a fine structure  of the antiresonances: each antiresonance splits     into a series of  narrow peaks, whose widths are governed  by dephasing. \cite{dmitriev}

The question which we address in this paper is the role of the spin-orbit (SO) and Zeeman interaction in the high-temperature  tunneling transport through the {\it ballistic single-channel }  ring with noninteracting electrons.

The effect of the SO interaction on the properties of  one-dimensional (1D) and quasi one-dimensional systems, in particular 1D quantum wires and rings,   has attracted  much attention.\cite{theorem,Stone1,history1,history6,history2,history3,history4,entin,Jap,devi,history5,exp1,exp2,citro2,plet,citro1,kov,cheng,Romeo,lobos,plet1,moldov,Aharony,dual,Michetti}    It is  known that in a 1D  noninteracting wire with an arbitrary spatial dependence of the spin-orbit coupling,
      the spin degree of freedom may be excluded  by a unitary transformation.\cite{entin}
            However, this is not the case  for
            a multiply-connected 1D system such as a  single-channel quantum ring. Though such a ring is actually a 1D system,
the interference between spin parts of  two counter-propagating electron waves    makes the problem less trivial.  Indeed,
the rotation of electron spin in the built-in SO  magnetic field  results in a spin phase shift  between clockwise and counterclockwise waves, which is a manifestation of the  Aharonov-Casher (AC) effect.\cite{AC,Stone}  The AC
 phase is the spin analog of the orbital AB phase. More precisely,  the AC phase   is additional with respect to  AB phase and exists even at zero external magnetic field ($\Phi=0$).   An  important consequence  is the existence of the AC  oscillations of zero-field conductance $G(0)$ with the strength of the SO coupling. The AC oscillations were intensively discussed theoretically \cite{Stone1,history1,history6,history2,history3,history4,history5,citro2,plet,citro1,kov,Romeo,lobos,plet1,moldov,Aharony,Michetti} and  their signatures  were  observed experimentally.\cite{exp1,exp2}
Another consequence, especially  important from the point of view of  possible applications,
is that
the unpolarized incoming electron beam acquires polarization after passing through the ring, so that the ring may serve as a spin polarizer.  The latter  effect was recently  discussed  in a number of publications \cite{history2,history3,history4, citro2, citro1,kov,Romeo,moldov,Aharony}  mostly concerned with the study of the  zero-temperature case.
The finite temperature effects were also analyzed on the basis of  numerical simulations. \cite{history4,citro1,moldov}

Here, we  develop an analytical theory of the spin-dependent transport through a ballistic single-channel  ring with two symmetric contacts     focusing on the high temperature case, $T\gg \Delta$. It will be shown  that  the spin-selective properties of the discussed setup survive thermal averaging.
 We will see that
  the SO interaction (in the absence of the Zeeman coupling) leads to  the splitting of  the high-temperature conductance antiresonances (Fig.~2)  
   into two dips (Fig.~4a),  the  distance between dips being proportional to the AC phase.   
   In the vicinities of the dips  the tunneling  electrons  acquire   polarization $\mathbf P(\phi) $
      [see Fig.~4b, Eqs.~\eqref{rho} and \eqref{Pvector}]. The vector  $\mathbf P(\phi) $ lies in the $(x,z)$ plane formed by two axes: the axis $x,$ which   connects  contacts $a$ and $b$ and the axial symmetry  axis of the ring  ($z$ axis perpendicular to the ring's plane).
      At zero external field ($\phi=0$) the  conductance $G(0)$ exhibits series of sharp AC antiresonances [see Eq.~\eqref{T_SO ACd}].       We also demonstrate that taking into  account the   Zeeman interaction leads to two effects (see Fig.~5): (i) emergence of additional  antiresonances  in  ${\cal T}(\phi)$  and   $\mathbf P(\phi)$; (ii) appearance  of nonzero polarization in $y$ direction.

\section{Ring with spinless electrons}
We start with a brief discussion (see also Ref.~\onlinecite{dmitriev})  of the  high-temperature conductance of the ring with   spinless electrons.   The purpose of this section is  to introduce  methods, which will be later generalized for the spinful  case.

\begin{figure}[ht!]
 \leavevmode \epsfxsize=5.0cm
 \centering{\epsfbox{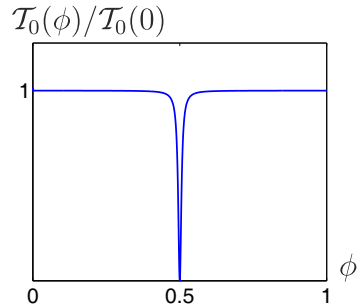}}
\caption{Dependence of the high-temperature  transmission coefficient on the magnetic flux 
in the spinless case for a ring with weak tunneling coupling to the leads.
\vspace{-3.5mm}}
\label{fig2}
\end{figure}

The conductance is calculated using the Landauer formula
\begin{equation}
G(\phi)=\frac{e^2}{2\pi\hbar} {\cal T}_0(\phi),
\label{Landauer}
\end{equation}
where
\begin{equation}
{\cal T}_0(\phi)= \langle{\cal T}_0(\phi,E) \rangle_E =-\int {\cal T}_0(\phi,E) \frac{\partial f}{\partial E}dE,
\label{average}
\end{equation}
is the energy-averaged value of the  transmission coefficient   ${\cal T}_0(\phi,E)$   and $f(E)$ is the Fermi-Dirac function.

We  consider both contacts to be identical and describe them with the following $S-$matrix:
\begin{equation}
S = \begin{bmatrix} t_r & t_{out} & t_{out} \\ t & t_b & t_{in} \\ t & t_{in} & t_b \end{bmatrix},
\label{S}
\end{equation}
which relates the amplitudes of three outgoing waves in the channels $(1',2',3')$ to the ones in the three incoming channels $(1,2,3)$ (see Fig.~3).

\begin{figure}[ht!]
 \leavevmode \epsfxsize=2.5cm \centering{\epsfbox{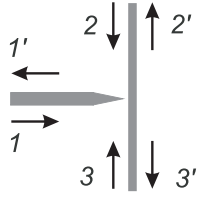}}
\caption{The scattering on contacts: the amplitude $t$ corresponds to processes $1\to 2'$ and $1 \to 3'$, $t_{out}$ - to $2\to 1'$ and $3\to 1'$, $t_r$ - to $1\to1'$,  $t_{in}$ - to $2\to 3'$ and $3 \to 2'$, $t_b$ - to $2\to 2'$ and $3 \to 3'.$
\vspace{-3.5mm}}
\label{fig3}
\end{figure}

In the simplest case of a point tunneling contact one can express all the elements of the $S$ matrix in terms of single real 
parameter $\gamma$:\cite{butt}
\begin{eqnarray}
t_{in} = \frac{1}{1+\gamma},\,\,\,t_b= -\frac{\gamma}{1+\gamma},\nonumber\\
t=t_{out} = \frac{\sqrt{2\gamma}}{1+\gamma},\,\,\, t_r = -\frac{1-\gamma}{1+\gamma}.
\label{t}
\end{eqnarray}
The case of weak tunneling coupling corresponds  to $\gamma \ll 1. $ The point metallic-like contact is described by $\gamma \sim 1. $

To find the transmission coefficient  we calculate the sum of the amplitudes of all the trajectories that correspond to electron passing through the ring from contact $a$ to contact $b.$ The summation of the amplitudes will be performed in the following way.
Each of the trajectories consists of the odd number $2n+1$ of semicircles  connecting the contacts.  The length of the trajectory is given by $L_n=\pi R (2n+1),$ where $R$ is the radius of the ring.   Let us denote the sum of the amplitudes of   all
trajectories with a given length $L_n$  (including trajectories with different number of backscatterings by contacts)
 as $\beta_n\exp(i k L_n),$ where $k=\sqrt{2mE}/\hbar$ is the electron wave vector.   The total transmission amplitude reads
 \be
 t_0(\phi,E)=\sum_{n=0}^{\infty} \beta_n\exp(i k L_n),
 \ee
so that the transmission coefficient  is given by
\begin{equation}
{\cal T}_0(\phi,E) =  |t_0(\phi,E)|^2=\sum\limits_{n=0}^{\infty}\sum\limits_{m=0}^{\infty}\beta_n  \beta_m^* e^{ik(L_n-L_m)}.\label{double_sum}
\end{equation}
  We  notice now that due to the condition $T\gg \Delta$  the  terms in Eq.~\eqref{double_sum}  corresponding to $n\neq m$  vanish after the averaging over  $E$ within the temperature window.   Hence, the  expression for the averaged  transmission coefficient becomes
\begin{equation}
{\cal T}_0(\phi) = \sum\limits_{n=0}^{\infty}|\beta_n|^2. \label{T0_sum}
\end{equation}

Next, we write $\beta_n = \beta_n^+ + \beta_n^-,$ where  $\beta_n^+$ ($\beta_n^-$) corresponds to trajectories ending with lower (upper) semicircle.
  It is easy to write the recurrence equations for $\beta_n^+$ and $\beta_n^-:$
\be
\begin{bmatrix} \beta_{n+1}^+\\ \beta_{n+1}^- \end{bmatrix}= \hat A_0 \begin{bmatrix} \beta_{n}^+\\ \beta_{n}^- \end{bmatrix},
\label{A0}
\ee
where the matrix $ \hat A_0=  \hat A_0(\phi)$ is given by
\be
\label{A00}
 \hat A_0  = \begin{bmatrix} t_{in}^2 e^{-2\pi i\phi} +t_b^2 & t_b t_{in} (e^{-2\pi i\phi} +1)\\ t_b t_{in} (e^{2\pi i\phi} +1) & t_{in}^2 e^{2\pi i\phi} +t_b^2 \end{bmatrix}.
\ee
Physically, the elements of matrix $\hat A_0$ are the amplitudes of four different trajectories of length $2\pi R$,  starting and ending on the contact $b.$

From Eqs.~\eqref{T0_sum} and \eqref{A0} we find
\bee
&&{\cal T}_0(\phi) = \sum\limits_{n=0}^{\infty}\,\left|\begin{bmatrix} 1\\1 \end{bmatrix}^\dagger \hat A_0^n\begin{bmatrix}  \beta_0^+\\ \beta_0^- \end{bmatrix}\right|^2\label{T0_sum2}\\
&&=\left(\begin{bmatrix} 1\\1 \end{bmatrix}\hspace{-1mm}\otimes\hspace{-1mm}\begin{bmatrix}  \beta_0^+\\ \beta_0^- \end{bmatrix}\right)^\dagger \hspace{-2mm}\frac{1}{1-\hat A_0\otimes \hat A_0^\dag}\begin{bmatrix} \beta_0^+\\ \beta_0^-  \end{bmatrix}\hspace{-1mm}\otimes\hspace{-1mm}\begin{bmatrix}  1\\1 \end{bmatrix},\nonumber
\eee
where $\otimes$ denotes the direct (Kronecker) product of matrices and \be \beta_{0}^+=t t_{out}e^{-i\pi\phi}, ~~~\beta_{0}^-=t t_{out}e^{i\pi\phi} \label{beta00} \ee  represent the amplitudes of shortest counterclockwise and clockwise   trajectories, respectively.

Using Eqs.~\eqref{t}, \eqref{A00}, \eqref{T0_sum2}, and  \eqref{beta00} after some algebra we get the following expression for the transmission coefficient:\cite{dmitriev}
\begin{equation}
{\cal T}_0(\phi) =\frac{2\gamma \cos^2\pi\phi}{\gamma^2+\cos^2\pi\phi}.\label{T0}
\end{equation}
In the almost closed ring with weak tunneling coupling,  $\gamma\ll 1,$
Eq.~\eqref{T0}  can be well-approximated with the  function
\begin{equation}
{\cal T}_0(\phi) =\frac{ 2\gamma\pi^2(\phi-1/2)^2}{\gamma^2+\pi^2(\phi-1/2)^2}\label{T00}
\end{equation}
[Eq.\eqref{T00}  is valid  for $0< \phi<1$].  This dependence is shown in Fig.~2. We see that there is a sharp antiresonance at $\phi = 1/2$ and ${\cal T}_0(1/2) = 0.$ The physics behind this behavior can be explained as follows.\cite{butt}
For each  trajectory there exists  a corresponding
   mirrored  (with respect to the $x$-axis) trajectory. The sum of the amplitudes of these  two trajectories is proportional to $e^{ikL_n} (e^{i(2|m|+1)\pi\phi}+e^{-i(2|m|+1)\pi\phi})$, where $m$ is a difference between the number of full clockwise and counterclockwise revolutions, $|m|\leq n.$ At $\phi=1/2$ this sum turns to zero for any $k$. Thus, the destructive interference at $\phi=1/2$ survives the thermal averaging.

\section{Ring with  spinful electrons}
In the spinful case the  Hamiltonian is given by
\be
\hat H = \hat H_{kin} + \hat H_Z + \hat H_{SO},
\label{Ham}
\ee
where
\be
\hat H_{kin} = - \frac{\hbar^2}{2mR^2}D_\varphi^2,
\label{kin}
\ee
is the kinetic energy, $D_\varphi = \partial/\partial\varphi + i\phi,$
\be
\hat H_Z = \frac12 \hbar \omega_Z\hat\sigma_z,
\label{ZM}
\ee
is the Zeeman term, $\hbar\omega_Z$ is the Zeeman splitting energy in the external field $\mathbf B $ parallel to $z$ axis,   and  $\hat H_{SO}$ corresponds to the SO coupling.

We assume that the SO interaction  is described by the Rashba Hamiltonian, which   for the  case of a straight wire looks
 $\hat H_{SO}=\alpha [\mathbf n \times \hat {\boldsymbol{\sigma}}]\, \mathbf p.$ Here $\mathbf n$ is the unit vector parallel to built-in electric field, $\hat {\boldsymbol{\sigma}}$ is the vector of the Pauli matrices, $\alpha$ is the constant of the SO interaction, and $\mathbf p$ is the  electron momentum. In a  curved wire, $\mathbf n$ depends on the coordinate,
 and the Hamiltonian becomes\cite{history1,history6,entin}
\begin{equation}
\hat H_{SO}=(\alpha/2) \{[\mathbf n \times \hat{ \boldsymbol{\sigma}}], \mathbf p\},
\label{Rashba}
\end{equation}
where $\{ \ldots\}$ stands for the anticommutator. For a ring with axially symmetric built-in field, $\mathbf n=(\cos\varphi \cos\theta,\sin\varphi\cos \theta,\sin \theta),$   we find from  Eq.~\eqref{Rashba}
 \be
\hat H_{SO} = -i\xi\frac{\hbar^2}{2mR^2}\left \{\begin{bmatrix} -\cos\theta && \sin\theta e^{-i\varphi} \\ \sin\theta e^{i\varphi} && \cos\theta \end{bmatrix}, D_\varphi\right\}.
\label{SO}
 \ee
 Here $\varphi$ is the angle coordinate of the electron in the ring, $\theta$ is the angle between 
 effective SO-induced magnetic field $\mathbf B_{{\rm eff}}$ (this field is proportional to $ \alpha [\mathbf p \times \mathbf n]$)  and the $z$ axis.
  The coefficient
$\xi = {\alpha m R}/{\hbar}$  entering Eq.~\eqref{SO} is the dimensionless parameter  characterizing the strength of SO interaction. Physically, $\xi$ is the angle of spin rotation  in the local field $\mathbf B_{{\rm eff}}$  during the time on the order of $R/v_F,$ where $v_F$ is the Fermi velocity. In the simplest case $\theta=0,$   $\xi$ is proportional to the angle of the spin rotation after passing around  the ring [see Eqs.\eqref{dd} and \eqref{rho}].

We study the problem  quasiclassically assuming that $k_FR\gg 1$ and $\alpha  \ll v_F$ (or, equivalently, $\xi\ll k_FR$). Within this approximation the combined  effect of the SO  and Zeeman interaction is fully described by the rotation of the electron spin in the  field $\mathbf B + \mathbf B_{{\rm eff}}, $ which varies along the electron trajectory. \cite{history1,history6}
Using Eqs.~\eqref{Ham}-\eqref{SO} we find (see Appendix A) the matrices of the spin rotation
\bee
&&\hat M_{a\to b}^\pm=\cos\pi \delta_\pm e^{i\vartheta_\pm \hat\sigma_y} \pm i\sin\pi \delta_\pm\hat\sigma_z, \label{M}\\
&&\hat M_{b \to a}^\pm = (\hat M_{a\to b}^\pm)^T.\nonumber
\eee
Here
$\hat M_{i\to j}^+ (\hat M_{i\to j}^-)$
describes  spin rotation for an electron passing  a semicircle from contact $i$ to contact $j$ with zero winding number in counterclockwise  (clockwise) directions,
 $\hat M^T$ denotes the transpose of a matrix $\hat M,$ and
\bee
&&\delta_\pm = |\varkappa_\pm| -\frac12, ~~~ e^{i\vartheta_\pm} = \frac{\varkappa_\pm}{|\varkappa_\pm|}, \label{kappa} \\
&&\varkappa_\pm =\frac12 + \xi e^{i\theta}  \mp\Omega_Z  . \nonumber
\eee
The strength of the Zeeman coupling is characterized by dimensionless parameter $\Omega_Z = {\omega_Z R }/{2v_F}.$ \cite{Fermi}

\begin{figure}[ht!]
 \leavevmode \epsfxsize=5.0cm
 \centering{\epsfbox{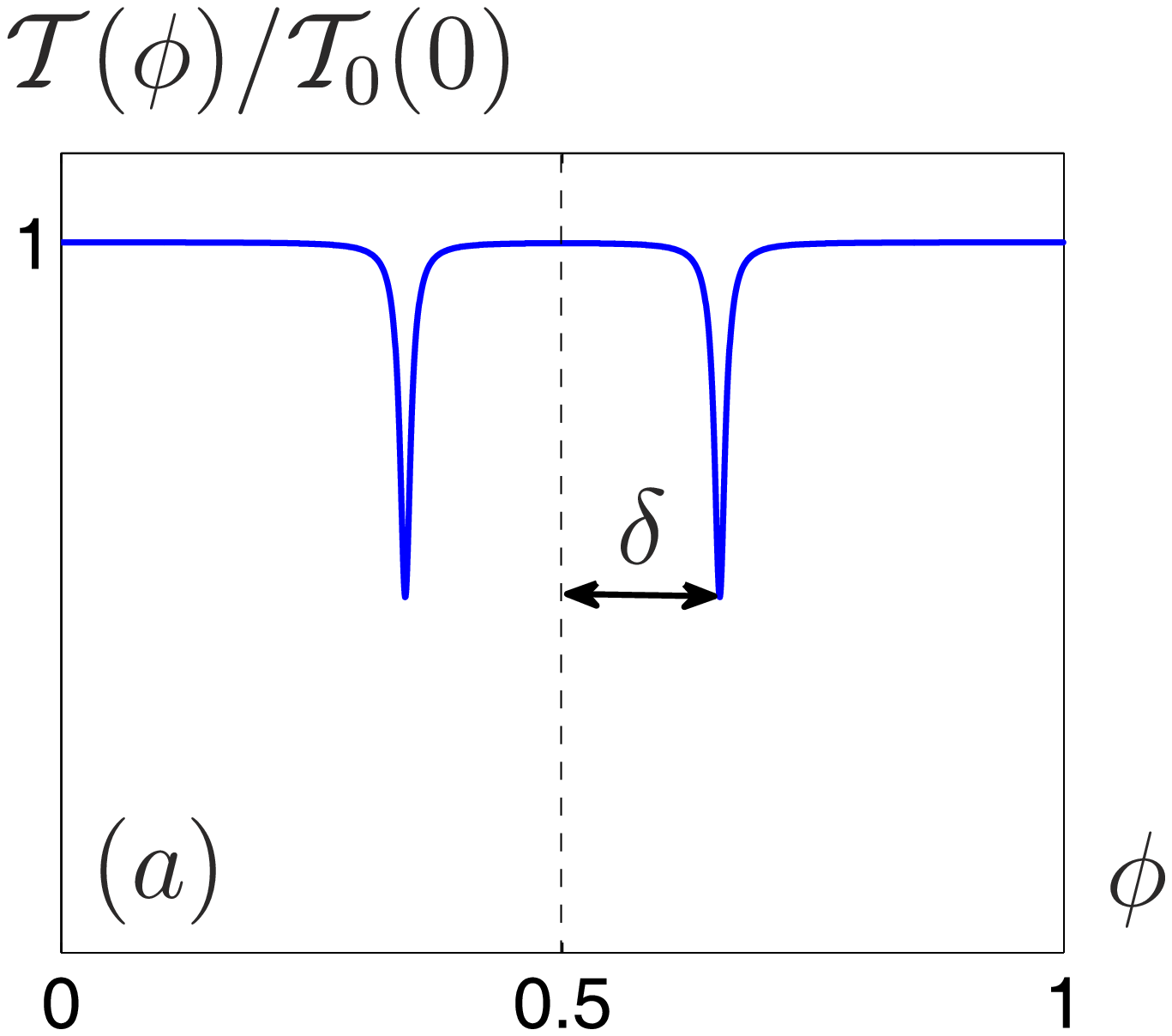}}
 \leavevmode \epsfxsize=5.0cm
 \centering{\epsfbox{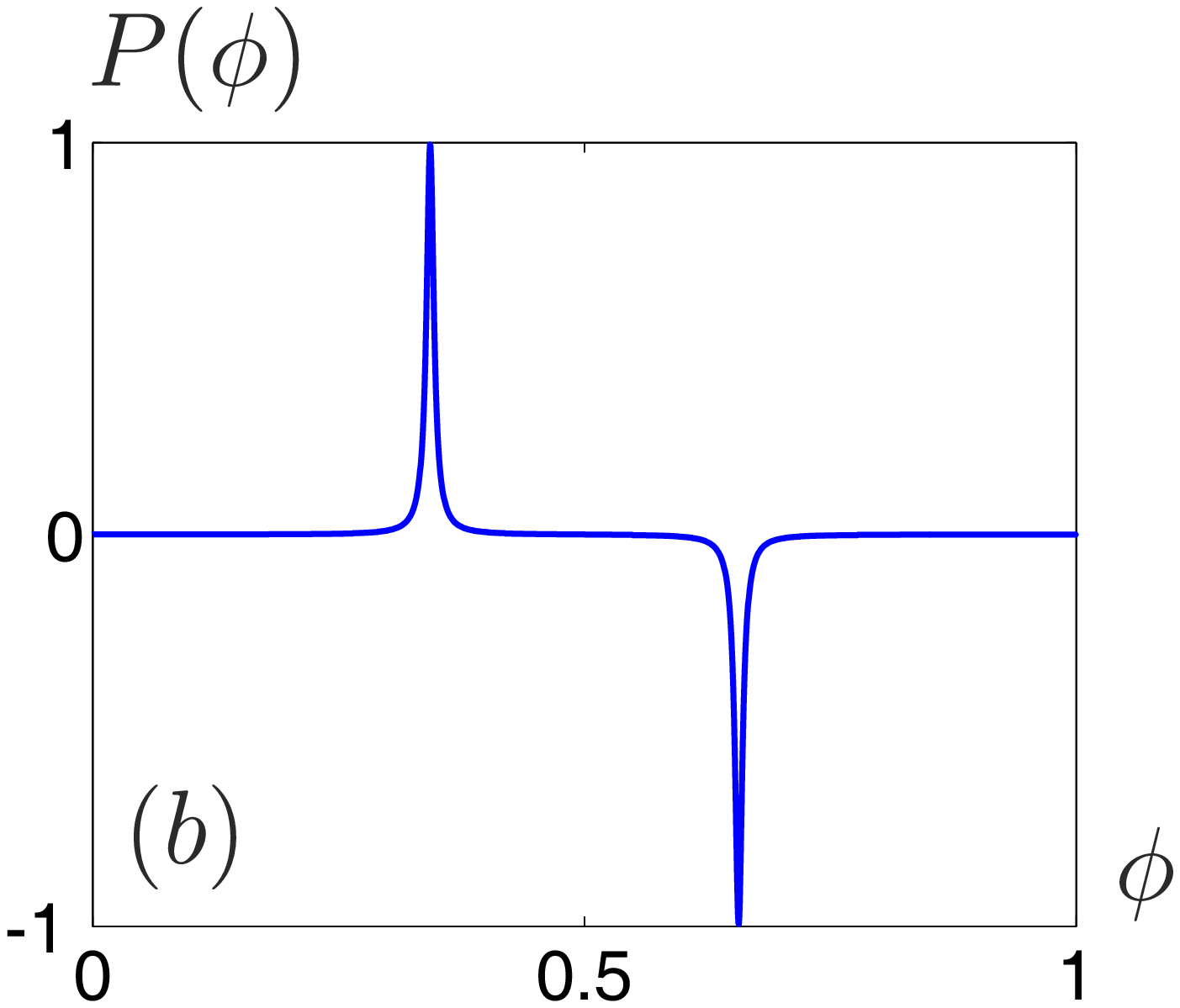}}
\caption{Transmission coefficient (a) and the
spin polarization of the transmitted electrons (b) in the direction of vector $\boldsymbol\rho$  for the ring with the SO interaction ($\gamma=0.02,~ \xi=0.2,~\theta=\pi/4,~ \Omega_Z=0$).
}
\label{fig4}
\end{figure}

The sum of the amplitudes of the   trajectories  having length $L_n,$ initial  spin state $|\chi_i\rangle,$ and final spin state $|\chi_f\rangle$ is given by $\langle \chi_f|\hat \beta_n|\chi_i\rangle,$ where $\hat \beta_n$ are now $2\times2$ matrices.
The amplitude of transmission through the ring with spin state changing from
$|\chi_i\rangle$ to $|\chi_f\rangle$ is given by $\langle\chi_i|\hat t|\chi_f\rangle$, where
\be \hat t(\phi,E)=\sum_{n=0}^\infty \hat \beta_n\exp(i k L_n). \ee
The transmission coefficient reads
\be {\cal T} = \frac12 \big\langle\mathrm{Tr}~{\hat t\, \hat t^\dagger}\big\rangle_E=\frac12\mathrm{Tr}~\hat{\cal T}, \ee
where
\be \hat{\cal T} = \sum_{n=0}^\infty \hat \beta_n \hat \beta_n^\dagger. \ee
The electrons passing through the ring acquire spin polarization. For the case of unpolarized incoming electron beam
 the spin polarization is calculated as  \be \mathbf P =\frac{\big\langle\mathrm{Tr}~{\hat{\boldsymbol\sigma}\hat t \hat t^\dagger}\big\rangle_E}{2{\cal T}}=
\frac{ \mathrm{Tr}~ \hat{\boldsymbol\sigma} \hat{\cal T}}{2{\cal T}},
\ee
and, therefore, is also expressed in terms of $\hat \beta_n.$

To find $\hat \beta_n$ we separate trajectories into two groups (just as in the previous section), writing $\hat \beta_n = \hat \beta_n^+ + \hat \beta_n^-,$
where $\hat \beta_n^+$ and $\hat \beta_n^-$ satisfy the recurrence equations, analogous to Eq.\eqref{A0}:
\begin{equation}
\begin{bmatrix} \hat \beta_{n+1}^+\\ \hat \beta_{n+1}^- \end{bmatrix}= \hat A \begin{bmatrix} \hat \beta_n^+\\ \hat \beta_n^- \end{bmatrix}.
\label{beta_rec}
\end{equation}
The   block  matrix $\hat A$  is given by
\be
\hat A \hspace{-1mm}= \hspace{-1mm}\begin{bmatrix}\hspace{-1mm} e^{-2\pi i\phi} t_{in}^2 \hat M_+  +t_b^2 \hat K &
 t_b t_{in}   (e^{-2\pi i\phi} \hspace{-0.5mm} \hat M_+  +\hspace{-1mm} \hat K)\\
 t_b t_{in}  (e^{2\pi i\phi} \hspace{-0.5mm}\hat M_-  + \hat K')&
e^{2\pi i\phi} t_{in}^2 \hat M_- +t_b^2 \hat K'   \end{bmatrix}.
\nonumber
\ee
Here
$ \hat M_+=\hat M_{a\to b}^+ \hat M_{b \to a}^+,~ \hat M_-=\hat M_{a \to b}^- \hat M_{b \to a}^- ,$ $
\hat K=\hat M_{a \to b}^+ \hat M_{b \to a}^-,~~ \hat K'=\hat M_{a \to b}^- \hat M_{b \to a}^+. $
The matrix $\hat M_+$ ($\hat M_-$) describes spin rotation after passing a full circle starting from contact $b$ and propagating in counterclockwise (clockwise) direction.
The matrix  $\hat K$ ($\hat K'$) is the spin rotation matrix for an electron, which  starts from the contact $b,$ then propagates along the lower (upper) shoulder of the interferometer and returns  back to  the contact $b$   after single backscattering on the contact $a.$

For the shortest trajectories we have
\bee
 &&\hat \beta_{0}^+= t t_{out} e^{-i\pi\phi}\hat M_{a\to b}^+,\label{beta_0} \\ \nonumber && \hat \beta_{0}^- = t t_{out} e^{i\pi\phi} \hat M_{a\to b}^-.
\eee
The matrix elements  of matrix $\hat {\cal T}$ can  be written in a form, analogous to Eq.\eqref{T0_sum2}:
\bee
&&\langle \chi_f| \hat {\cal T}|\chi_i\rangle  =     \sum_{n,k}  \langle \chi_f| \hat \beta_n |\chi_k\rangle \langle\chi_k|\hat \beta_n^\dagger|\chi_i\rangle \label{TTkron}\\
&&=\sum_k\hspace{-1mm}
\begin{bmatrix}\chi_f\\\chi_f \end{bmatrix}^\dagger\hspace{-2mm}\otimes\hspace{-1mm}\begin{bmatrix}  \hat \beta_0^+\chi_k\\ \hat \beta_0^-\chi_k \end{bmatrix}
^\dagger \hspace{-2mm}\frac{1}{1\hspace{-1mm}-\hspace{-1mm}\hat A\otimes \hat A^\dag}\hspace{-1mm}\begin{bmatrix} \hat \beta_0^+\chi_k\\ \hat \beta_0^-\chi_k  \end{bmatrix}\hspace{-1mm}\otimes\hspace{-1mm}\begin{bmatrix}   \chi_i\\ \chi_i \end{bmatrix}.\nonumber
\eee

In the following sections we will use Eqs.~ \eqref{M}-\eqref{TTkron} to calculate the full transmission coefficient and the spin polarization for a ring with the SO and Zeeman interactions.

\subsection{Ring with SO interaction (zero Zeeman coupling)}
In the absence of the Zeeman coupling   ($\Omega_Z=0$)
we find from Eq.~\eqref{kappa}
\be \delta_+ = \delta_- = \delta,\,~~~ \vartheta_+ = \vartheta_- = \vartheta. \label{dt}\ee
Here
\bee && \delta =\sqrt{\frac14 +\xi \cos\theta +\xi^2}-\frac12,  \label{dd}\\
&& \tan \vartheta =\frac{\xi \sin\theta}{1/2+\xi\cos\theta}. \label{tt}\eee
 Now we will make use of the important feature of SO interaction: if the electron travels along a certain trajectory and then returns
 to the initial point along the same trajectory moving in the opposite direction, its spin returns to the original state.  This implies that
  \be \hat K=\hat K'= 1,~~ \hat M_+=\hat M_-^{-1} = \hat M  \label{MMM}\ee
    [one can easily check Eq.~\eqref{MMM} using Eqs.~\eqref{M} and \eqref{dt}].
    These properties
    essentially simplify further calculations.  The block  matrix $\hat A$  is now  fully expressed in terms of matrix $\hat M$
 \be
\hat A = \begin{bmatrix} t_{in}^2 e^{-2\pi i\phi} \hat M  +t_b^2  &
 t_b t_{in}   (e^{-2\pi i\phi} \hat M  + 1)\\
 t_b t_{in}  (e^{2\pi i\phi} \hat M^{-1}  +1)&
t_{in}^2 e^{2\pi i\phi} \hat M^{-1} +t_b^2  \nonumber \end{bmatrix}.
\ee
   Next, we write \be \hat M =\exp(-i\boldsymbol\rho\hat{\boldsymbol\sigma}/2),\ee
    where $\boldsymbol\rho$ is the  vector of spin rotation for counterclockwise propagation around the ring (starting from contact $b$). From Eqs.~\eqref{M}, \eqref{dt}, \eqref{dd}, and \eqref{tt} we find
 \be
 \boldsymbol\rho = 4\pi \delta(\boldsymbol e_x \sin \vartheta  - \boldsymbol e_z \cos \vartheta  ).
 \label{rho} \ee
 The eigenvectors of $\hat M$ are the spinors $\chi^{\uparrow}$ and $\chi^{\downarrow}$ corresponding to spin orientation along $\boldsymbol\rho$ and $-\boldsymbol\rho:$
\bee && \hat M \chi^{\uparrow} =  \exp(- i 2\pi |\delta|) \chi^{\uparrow}, \label{up}\\
&& \hat M \chi^{\downarrow} =  \exp( i 2\pi |\delta|) \chi^{\downarrow}. \label{down}
 \eee

As follows from Eqs.~\eqref{up} and \eqref{down}, $  2\pi |\delta|$ is the Aharonov-Casher (AC)  phase  \cite{AC,Stone} induced by the SO interaction.
Using Eq.~\eqref{dd} we find that  $\delta$
     lies in the interval between $\sqrt{\xi^2+1/4}-1/2$  and  $\xi.$  These limiting values are realized, respectively,   for   $\theta=\pi/2$  ($\mathbf B_{{\rm eff}}$ is parallel to the ring plane) and $\theta=0$ ($\mathbf B_{{\rm eff}}$  is parallel to the $z$ axis).

For $\xi \gg 1,$ the frequency of  spin precession in the field $\mathbf B_{{\rm eff}}$ is much larger than the   orbital frequency $v_F/R$ and the direction of the spin follows adiabatically the direction of $\mathbf B_{{\rm eff}}.$
In this case Eq.~\eqref{dd} simplifies
\be 2\pi \delta = 2\pi \xi - \pi (1-\cos \theta), ~~\text{ for}~~ \xi \gg 1     \label{berry}.\ee
 Thus, in the adiabatic limit  the AC phase separates into  two parts:\cite{history1} dynamical contribution $2 \pi\xi$ and geometrical SO  Berry phase \cite{berry}  $\pi (1-\cos \theta)$ which is the half of the solid angle subtended by $\mathbf B_{{\rm eff}}$ when electron passes the full circle. \cite{comment2}

In order to find the recurrence equations for $\hat \beta_n^\pm$ we  now introduce the spinors $\tilde\chi^{\uparrow\downarrow} = \exp(\mp i\pi|\delta|)(\hat M_{a\to b}^+)^{-1}\chi^{\uparrow\downarrow},$ which  are transformed to $\chi^{\uparrow\downarrow}$ when the electron propagates from contact $a$ to contact $b$ [the phase multiplier $\exp(\mp i\pi|\delta|)$ is added for convenience].
Using Eq.\eqref{beta_rec} we get $\langle\chi^{\downarrow}|\hat\beta_n|\tilde\chi^{\uparrow}\rangle=\langle\chi^{\uparrow}|\hat\beta_n|\tilde\chi^{\downarrow}\rangle=0$, so that $\hat\beta_n$ can be written as
\be
\hat \beta_n= \beta_n^\uparrow |\chi_\uparrow \rangle \langle \tilde \chi_\uparrow| + \beta_n^\downarrow|\chi_\downarrow \rangle \langle \tilde \chi_\downarrow|.
\label{t_phi_e}
\ee
 For $\beta_n^{\uparrow} = \langle\chi^{\uparrow}|\hat\beta_n|\tilde\chi^{\uparrow}\rangle$ we get the recurrence equations
\begin{equation}
\begin{bmatrix} \beta_{n+1}^{\uparrow+}\\ \beta_{n+1}^{\uparrow-} \end{bmatrix}=
\hat A_0(\phi+|\delta|) \begin{bmatrix} \beta_n^{\uparrow+}\\ \beta_n^{\uparrow-} \end{bmatrix}, \label{beta_chi}
\end{equation}
Here $\hat A_0(\phi)$ is given by Eq.~\eqref{A00} and
\begin{equation}
\begin{bmatrix}\beta_0^{\uparrow+}\\\beta_0^{\uparrow-} \end{bmatrix} = \begin{bmatrix} e^{-i\pi(\phi+|\delta|)}\\ e^{i\pi(\phi+|\delta|)}\end{bmatrix}\label{beta_chi0}.
\end{equation}
We see that the quantities $\beta_n^{\uparrow\pm}$ satisfy the same recurrence equations as the ones in the spinless case  [see Eq.~\eqref{A0}] with the replacement  $\phi$  with  $\phi +|\delta|.$  One can easily show that the recurrence equations for  $\beta_n^{\downarrow}=\langle\chi^{\downarrow}|\hat\beta_n|\tilde\chi^{\downarrow}\rangle$  are given by Eqs.~\eqref{beta_chi} and \eqref{beta_chi0} with the replacement $\phi +|\delta|$ with $\phi -|\delta|.$
As a result, we find
\be
\hat{\cal T}(\phi) = {\cal T}_0(\phi-|\delta|)|\chi_\uparrow \rangle \langle \chi_\uparrow|+{\cal T}_0(\phi+|\delta|)|\chi_\downarrow \rangle \langle \chi_\downarrow|,
\ee
 where ${\cal T}_0$ is the transmission coefficient of the spinless electrons given by Eq.(9).
The expressions for the  full transmission coefficient  and the spin polarization become
\bee
 &&{\cal T}(\phi) = \frac{ {\cal T}_0(\phi+\delta)+ {\cal T}_0(\phi-\delta) }{2},\label{T_SO}\\
 &&\mathbf P(\phi)= P(\phi)\frac{\boldsymbol{\rho}}{\rho},\label{Pvector}
 \eee
where
 \be
 P(\phi)
=\frac{{\cal T}_0(\phi+|\delta|)- {\cal T}_0(\phi-|\delta|)}{{\cal T}_0(\phi+|\delta|)+ {\cal T}_0(\phi-|\delta|)}.\label{S_SO}
\ee
It is worth noting that Eq.~\eqref{T_SO} is in agreement with the general theorem, relating any transport property of 1D system with the SO interaction with the same property without the SO interaction.\cite{theorem}

The dependencies of the conductance and the spin polarization on magnetic flux are schematically depicted in Fig.4. As seen, there are two dips (per period) in the function ${\cal T}(\phi)$, corresponding to $\phi = 1/2 \pm \delta +N.$
 At these two points the incoming electrons with spin states described, respectively, by $\tilde\chi^{\downarrow}$ and $\tilde\chi^{\uparrow}$ are totally blocked by the destructive interference.  Therefore, the tunneling current becomes fully polarized in the direction of $\boldsymbol\rho$ for $\phi = 1/2 \pm \delta + N.$

 We see that SO-induced splitting of the  resonances is proportional to the AC phase $2\pi\delta$.    Eqs.~\eqref{T_SO} and \eqref{S_SO}
 reveal coexisting of  two types of oscillations: the AB oscillations with $\phi$ and $AC$ oscillations with $\delta.$
  Importantly, AC oscillations  of tunneling conductance exist even in the case of zero external field.  Indeed, for $\phi=0$,    we have
    \be
 {\cal T} = {\cal T}_0(\delta), ~~~
   P  = 0 .\label{T_SO ACd}
\ee
 Here we took into account that ${\cal T}_0(\delta)$ is an even function. Thus,  transmission coefficient exhibits  the AC oscillations with the period $\delta=1$. For the case of almost closed ring, $\gamma \ll 1,$ the oscillations have the form of sharp    antiresonances periodic in  $\delta$.

 In conclusion of this section, we note that Eqs.~\eqref{T_SO},  \eqref{S_SO}, and   \eqref{T_SO ACd} are valid for $T\gg\Delta$ and arbitrary strength of tunneling coupling ($0<\gamma <\infty$).  They  represent a generalization   of the analytical results obtained previously \cite{history2,history4,history5,citro2,citro1} for $T=0$ and strong  tunneling coupling  (metallic-like contacts, $\gamma \simeq 1$).

\subsection
{Interplay of    spin-orbit and Zeeman interactions}

Next we discuss the role  of  the Zeeman interaction. Taking this interaction into account requires much more tricky calculations.
The point is that the properties \eqref{MMM} are no longer valid when the time reversal symmetry is broken. Consequently, the elements of block matrix $\hat A$ can not be expressed in terms of a single rotation matrix.
In principle, the  expressions for ${\cal T}$ and $\mathbf P$ can be derived from \eqref{TTkron},
where both averaging over the temperature window and summation over winding number $n$ are already done.
This equation, indeed, turns out to be very useful for numerical simulations.
However, the analytical expressions  obtained with the use of \eqref{TTkron}  for the case of arbitrary $\gamma$ turn out to 
be very cumbersome and we do not present them here.
We restrict ourselves with the analytical study of the almost closed ring, $\gamma\ll1. $  For this case, the calculations presented in the  Appendix B yield
\begin{widetext}
\bee
&&{\cal T}(\phi) = \frac{c_-^2\left[{\cal T}_0\left(\phi +\delta\right)+ {\cal T}_0\left(\phi -\delta\right)\right]+ s_-^2\left[{\cal T}_0\left(\phi +\delta^\prime\right)+ {\cal T}_0\left(\phi -\delta^\prime\right)\right]}{2},\label{T_SOZ}\\
&&P_x(\phi) = \frac{ s_+ c_-\left[{\cal T}_0\left(\phi +\delta\right)- {\cal T}_0\left(\phi -\delta\right)\right]
+ s_- c_+ \left[{\cal T}_0\left(\phi +\delta^\prime\right)- {\cal T}_0\left(\phi -\delta^\prime\right)\right]}{2{\cal T}(\phi)},\label{S_SOX}
\\
&&P_y(\phi) = \frac{s_-c_-
\left[{\cal T}_0^\prime\left(\phi +\delta^\prime\right)+ {\cal T}_0^\prime\left(\phi -\delta^\prime\right)
-{\cal T}_0^\prime\left(\phi +\delta\right)- {\cal T}_0^\prime\left(\phi -\delta\right)\right]}{{2\cal T}(\phi)} ,\label{S_SOZ}
\\
&&P_z(\phi) = \frac{ c_+ c_-\left[{\cal T}_0\left(\phi -\delta\right)- {\cal T}_0\left(\phi +\delta\right)\right]
 +s_+s_- \left[{\cal T}_0\left(\phi +\delta^\prime\right)- {\cal T}_0\left(\phi -\delta^\prime\right)\right]}{2{\cal T}(\phi)},
\label{S_SOZ}
\eee
\end{widetext}
where
\bee
&&\delta = \frac{\delta_++\delta_-}{2},\,\, \delta^\prime = \frac{\delta_+-\delta_-}2+\frac12,\label{ddprime} \\
&&s_\pm = \sin\left(\frac{\vartheta_+\pm\vartheta_-}{2}\right),\label{spm}\\
&&c_\pm = \cos\left(\frac{\vartheta_+\pm\vartheta_-}{2}\right)\label{cpm},
\eee
and
\be
{\cal T}_0^\prime(\phi) = \frac{2\pi\gamma^2(\phi - 1/2)}{\gamma^2+\pi^2(\phi-1/2)^2}.\label{Tprime}
\ee
These results are shown in Fig.~5. We see that in the presence of  the Zeeman interaction instead of the two antiresonances there are four  ones (per period)  corresponding to the flux values $1/2 \pm \delta +N$ and $1/2 \pm \delta^\prime + N.$ In the vicinity of each antiresonance the outgoing  electrons are polarized.  Importantly,  the Zeeman coupling induces  nonzero polarization in $y$ direction.  One may notice that   the dependence  $P_y(\phi)$ is qualitatively different from dependencies  $P_x(\phi)$ and $P_z(\phi).$ First of all, the    peaks in $P_y(\phi)$ are asymmetric [see Eq.~\eqref{Tprime}]  in contrast to the peaks in $P_x(\phi)$ and $P_z(\phi).$  Secondly, all four resonances in $P_y(\phi)$ have the same amplitudes.
\begin{figure}[ht!]
 \leavevmode \epsfxsize=5.0cm
 \centering{\epsfbox{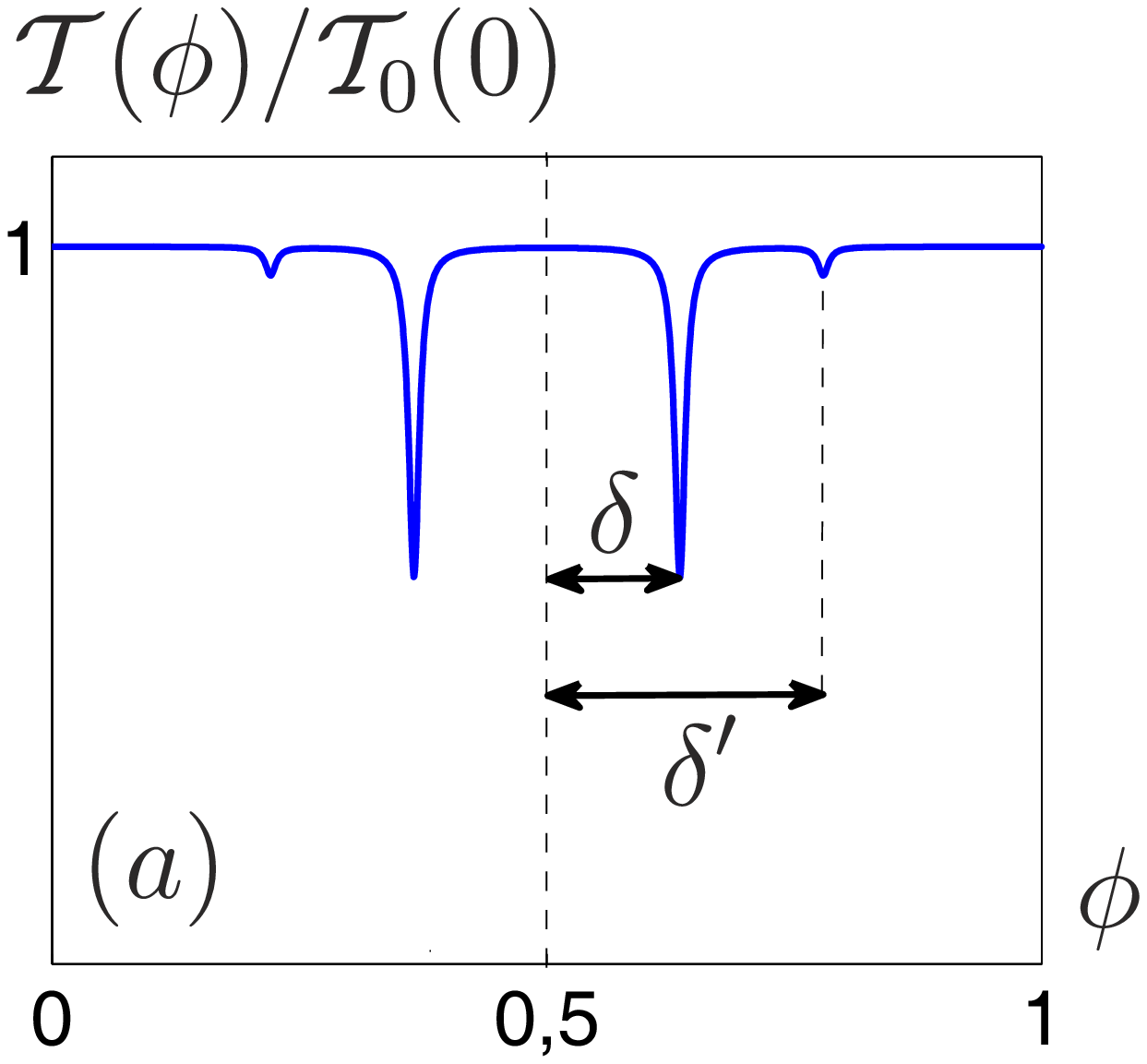}}
 \leavevmode \epsfxsize=5.0cm
 \centering{\epsfbox{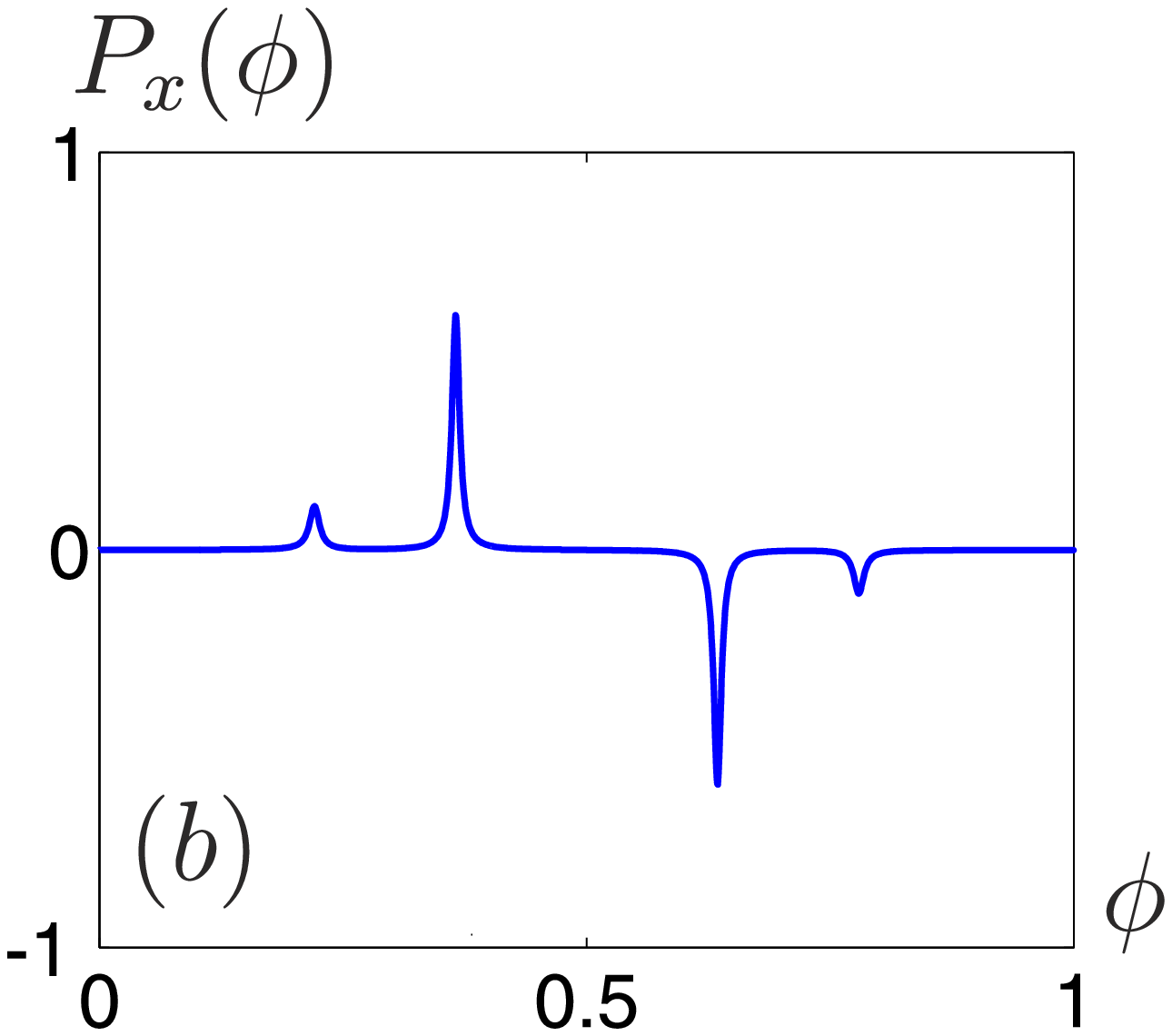}}
 \leavevmode \epsfxsize=5.0cm
 \centering{\epsfbox{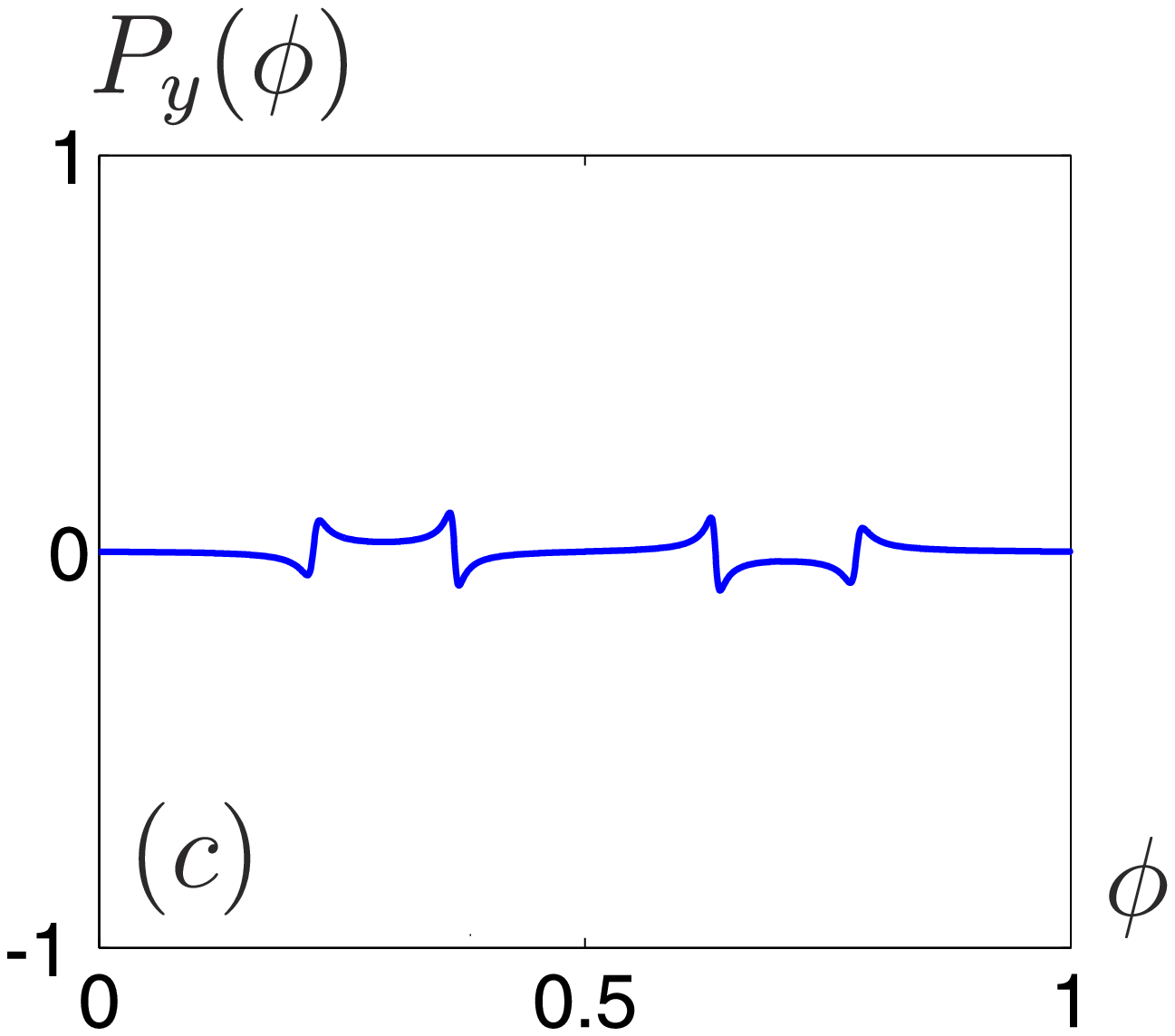}}
 \leavevmode \epsfxsize=5.0cm
 \centering{\epsfbox{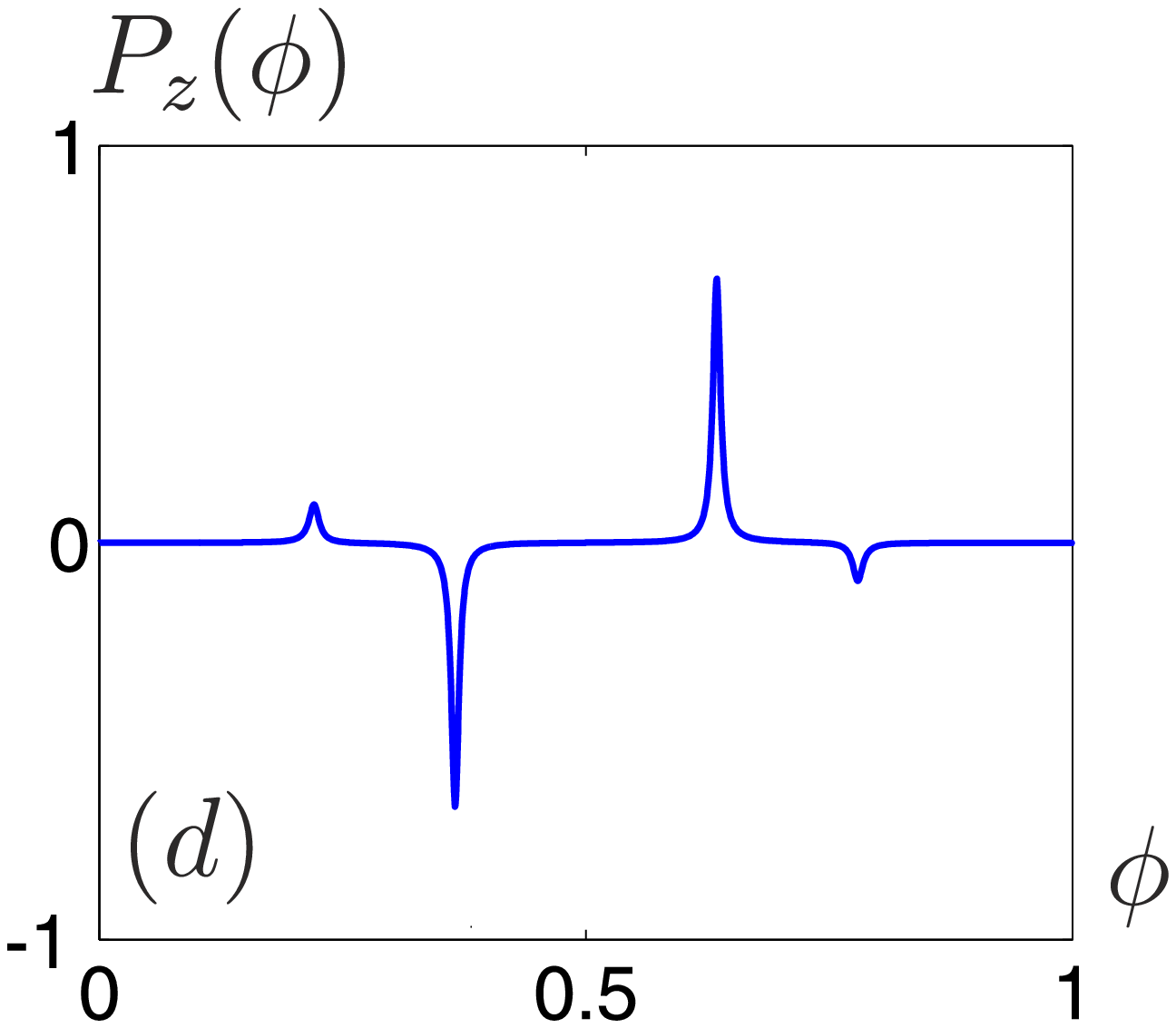}}
\caption{ Transmission coefficient through the ring with the SO and Zeeman interactions (a) and spin polarization of the transmitted electrons in the $x$, $y$ and $z$ directions (b), (c) and (d), respectively ($\gamma=0.02,~ \xi=1.9,~ \theta=\pi/4,~ \Omega_Z=1$).}
\label{fig5}
\end{figure}

Next, we consider some limiting cases. For weak Zeeman coupling, $|\Omega_Z|\ll\max\{1,|\xi|\},$ one finds
\bee
&&\delta \approx \frac{\xi\cos\theta +\xi^2}{\sqrt{1/4 +\xi\cos\theta+\xi^2}+1/2}, ~\delta^\prime \approx 1/2-\Omega_Z, \nonumber \\
&&c_- \approx 1,~~s_-\approx \frac{\Omega_Z\xi\sin\theta}{1/4+\xi\cos\theta+\xi^2},\\%\nonumber\\
&&c_+ \approx \frac{1/2+\xi\cos\theta}{\sqrt{1/4 +\xi\cos\theta+\xi^2}},\nonumber\\
&&s_+ \approx \frac{\xi\sin\theta}{\sqrt{1/4 +\xi\cos\theta+\xi^2}}.\nonumber
\eee

In the strong Zeeman coupling limit, $|\Omega_Z|\gg\max\{1,|\xi|\},$ we obtain
\bee
&&\delta \approx 1/2-\Omega_Z ,~~ \delta^\prime \approx \xi\cos\theta, \nonumber\\
&& c_-\approx \frac{\xi\sin\theta}{|\Omega_Z|},~~s_- \approx \mathrm{sign}~\Omega_Z, \\
&& c_+ \approx \frac{\xi\sin\theta\left(\frac12 +\xi\cos\theta\right)}{\Omega_Z^2}, ~~ s_+ \approx 1.\nonumber
\eee

We see that in  both limiting cases  there are two deep antiresonances, the positions of which are controlled by the strength of the SO interaction, and two small ones with the positions controlled by the Zeeman coupling.

The competition between spin-orbit and Zeeman coupling is clearly seen in the case  $\xi\gg 1, ~~\Omega_Z \gg 1,$   and arbitrary relation between $\xi$ and  $\Omega_Z .$ In particular, the amplitudes of peaks in ${\cal T}(\phi)$ are given by
\bee
&&\frac{s_-^2}{2}  \approx \frac12 \left(1+\frac{\Omega_Z^2-\xi^2}{\sqrt{(\Omega_Z^2-\xi^2)^2+4\Omega_Z^2\xi^2\sin^2\theta}}\right),\nonumber\\
&&\frac{c_-^2}{2}\approx \frac12 \left(1-\frac{\Omega_Z^2-\xi^2}{\sqrt{(\Omega_Z^2-\xi^2)^2+4\Omega_Z^2\xi^2\sin^2\theta}}\right).\nonumber
\eee
As follows from these equations, for $\xi\approx\Omega_Z$ all four antiresonances have the same amplitudes.

The dependencies of the positions of four antiresonances on $\Omega_Z$ with fixed $\xi$ is shown in  Fig.~6.   It is noteworthy that in the special case $\theta =\pi/2$, the distance between the deep antiresonances tends to zero when the strength of the Zeeman coupling increases, so that ${\cal T}(\phi)\to{\cal T}_0(\phi)$ when $\Omega_Z \to \infty.$ This is also illustrated  in Fig.~7 where  ${\cal T}(\phi)$ is plotted for  $\theta=\pi/4$ and $\theta=\pi/2.$

\begin{figure}[ht!]\label{fig6}
 \leavevmode \epsfxsize=5.0cm
 \centering{\epsfbox{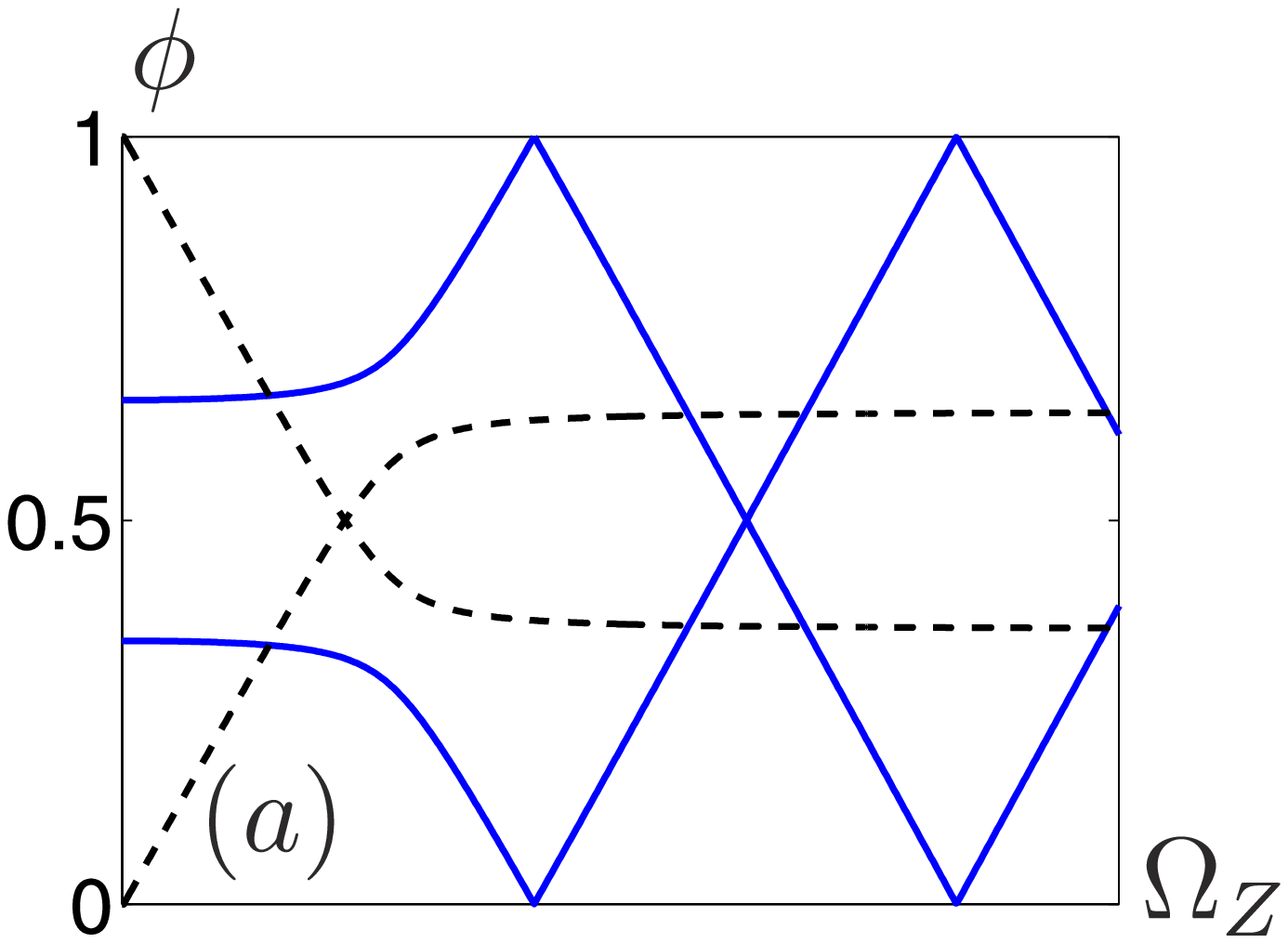}}
\leavevmode \epsfxsize=5.0cm
 \centering{\epsfbox{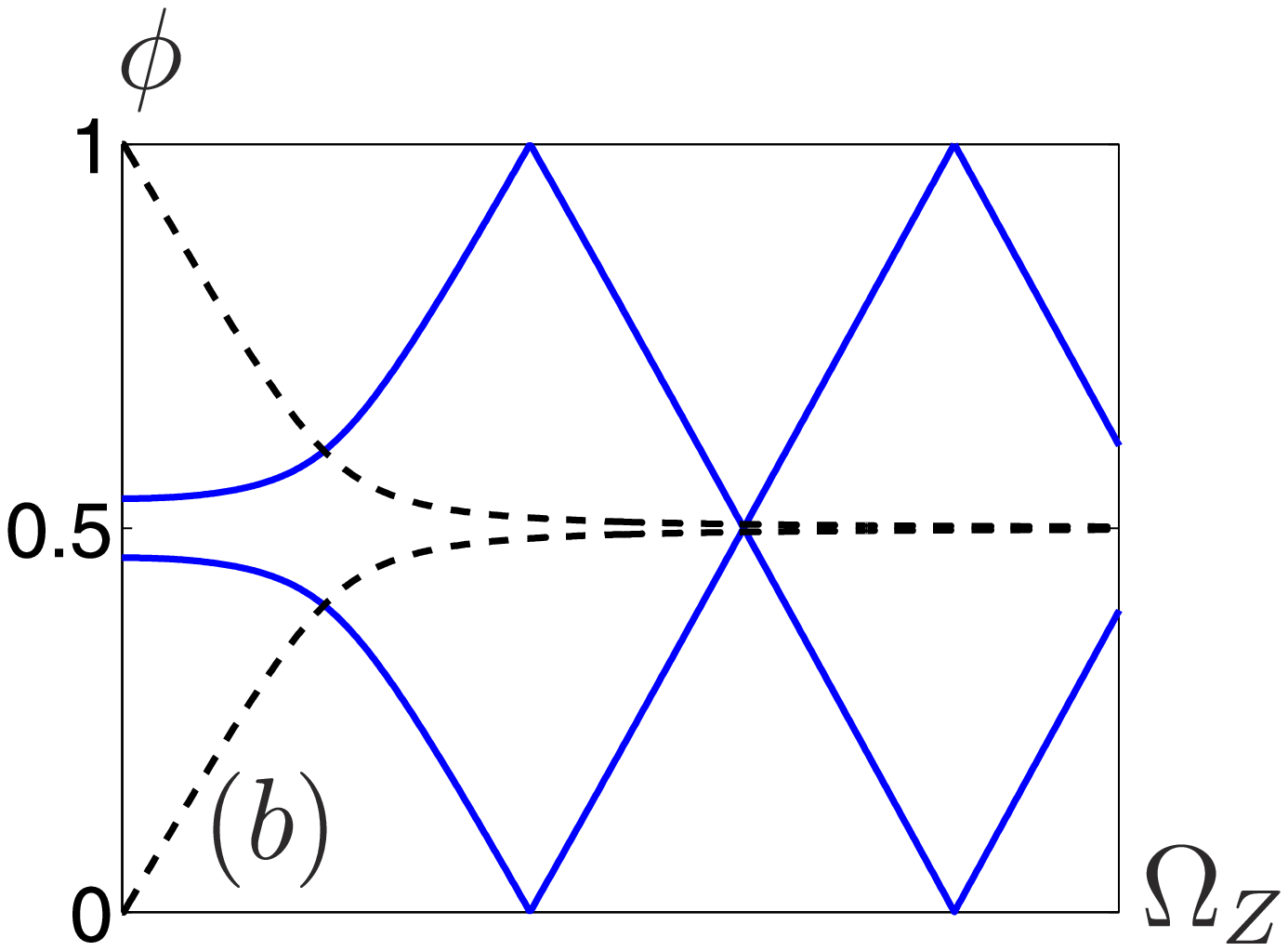}}
\caption{Positions of antiresonances in the region $0<\phi<1$ as  functions of $\Omega_Z$ for $\xi=0.2$ and different values of $\theta$:  $\theta=\pi/4$ (a) and  $\theta=\pi/2$ (b). Solid and dashed lines correspond to $ 1/2 \pm \delta + N$ and  $ 1/2 \pm \delta^\prime + N,$ respectively. }
\end{figure}
\begin{figure}[ht!]\label{fig7}
 \leavevmode \epsfxsize=5.0cm
 \centering{\epsfbox{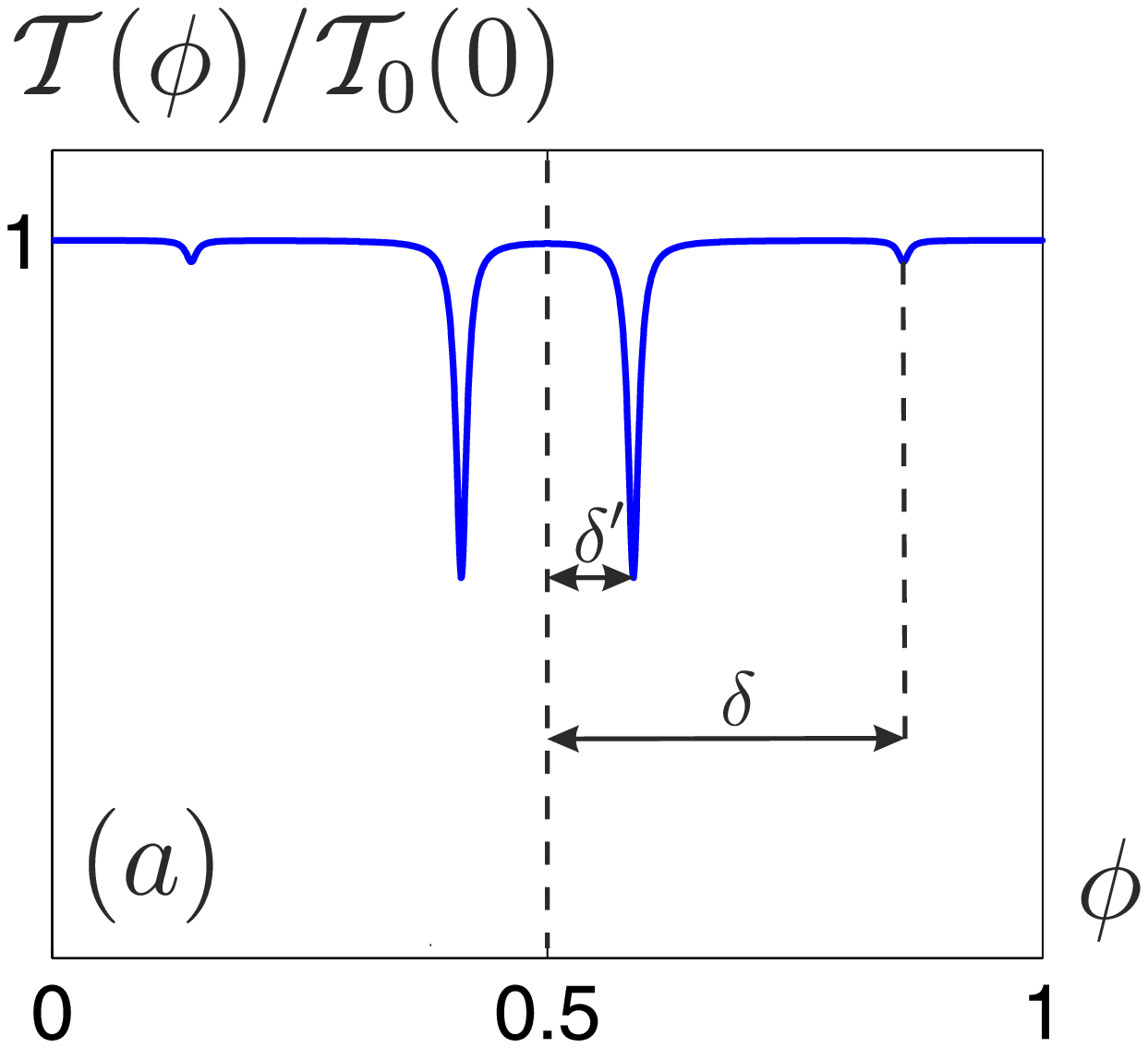}}
\leavevmode \epsfxsize=5.0cm
 \centering{\epsfbox{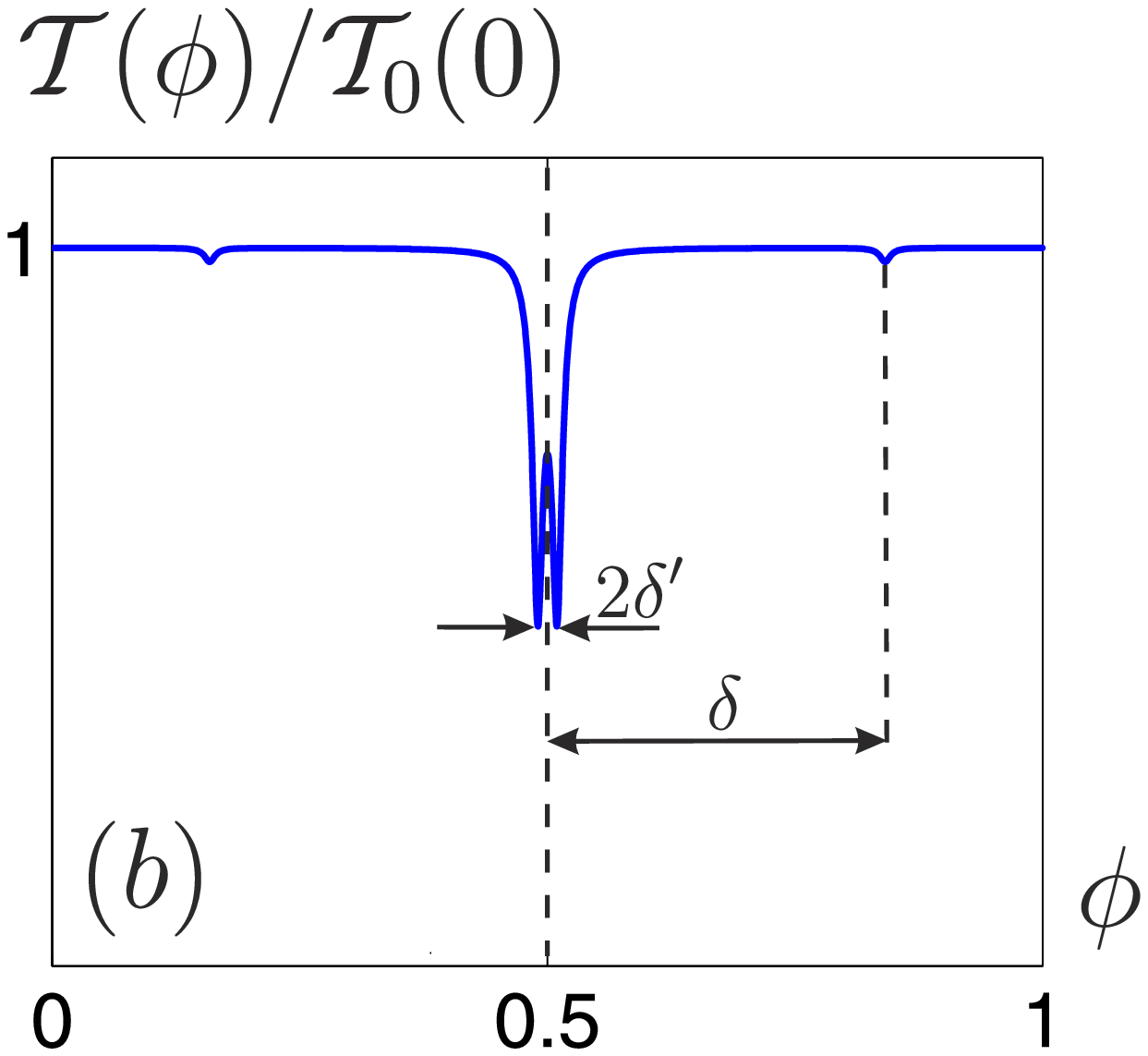}}
\caption{Transmission coefficient  for the  strong Zeeman splitting: $\xi=1.6,~\Omega_Z=5,~\theta=\pi/4$ (a),  $\xi=1.6,~\Omega_Z=8,~\theta=\pi/2$ (b). For $\theta = \pi/2$ the effect of the SO coupling is suppressed with  increasing the Zeeman coupling (b).}
\end{figure}

At the end of this section we note that  the Zeeman interaction may also lead to inhomogeneous broadening of the antiresonances. Indeed, in the above calculations we replaced in the dimensionless parameter    $\Omega_Z$   the energy-dependent electron velocity $v$ with $v_F.$ In fact, the positions of antiresonances [see Eq.~\eqref{ddprime}] depend on $v,$ so that one should average Eqs~\eqref{T_SOZ}-\eqref{S_SOZ} over the temperature fluctuations of $\Omega_Z.$ Having in mind that such fluctuations are  on the order of $\Omega_Z T/E_F,$ one can conclude that  the inhomogeneous broadening should be taken into account when $\Omega_Z T/E_F$ becomes larger than $\gamma.$

\section{Summary}

In this paper we calculated the  high-temperature  transmission coefficient ${\cal T}(\phi)$  and the spin polarization $\mathbf P(\phi)$ of the electrons  tunneling through  a  single-channel ring with the Rashba  SO interaction threaded by magnetic flux.  We obtained
analytical expressions for ${\cal T}(\phi)$ and $\mathbf P(\phi)$ valid  for arbitrary strength of the tunneling coupling.  Both ${\cal T}(\phi)$ and $\mathbf P(\phi)$ reveal
coexistence of two types of  periodic   oscillations:
the Aharonov-Bohm oscillations with magnetic flux and the Aharonov-Casher oscillations with the strength of  SO interaction.
For weak tunneling coupling,  the oscillations have the form of the sharp antiresonances periodic in $\phi$ and $\delta.$
Specifically, in the absence of the Zeeman coupling there are two antiresonances (per period) in the dependence  ${\cal T}(\phi)$ (instead of one antiresonance  in the spinless case). In the vicinity of each   antiresonance, the electron beam passing through the ring acquires strong spin
polarization  directed in the $(x,z)$ plane formed by axial symmetry axis of the ring ($z$ axis) and the line connecting two contacts.
We also discussed  the  influence of the Zeeman interaction on the interference picture and showed that two additional  antiresonances appear both in    ${\cal T}(\phi)$ and $\mathbf P(\phi).$ Also, the Zeeman coupling leads to appearance of nonzero spin polarization in $y$ direction.

Let us finally briefly discuss some unsolved problems.
In the above calculations we assumed that electrons are noninteracting.        The detailed analysis  of the effects caused by electron-electron interaction is out of scope of the current paper.
Here we restrict ourselves to a very  brief qualitative discussion (for the case $T\ll \Delta,$ the role of the electron-electron interaction in a ring with SO interaction
was discussed in Refs.~\onlinecite{plet}).  First of all, the interaction leads to a renormalization of the tunneling rate due to  Luttinger-liquid correlations specific for purely 1D systems.
This implies that
in all equations derived above one should replace $\gamma$ by renormalized tunneling rate $\tilde \gamma$ (for discussion of renormalization of
$3\times3$ matrix of the tunneling contact in the spinless case see Refs.~\onlinecite{rao1,rao2,aristov,aristov1,aristov2}).
  This does not change our results much, since $\gamma$ is a phenomenological parameter of the model.   What is more important is that the interaction suppresses the interference between clockwise and counterclockwise propagating electron waves, thus leading to a broadening of the  resonances.
It is expected \cite{dmitriev1}  that this broadening is  negligible when interaction is so weak that
\be \tilde \gamma \Delta \gg \Gamma_\varphi^0,  \label{ineq}\ee
where $\Gamma_\varphi^0 \simeq g T $ is the electron-electron scattering  rate in the infinite single-channel wire \cite{yashenkin} (here $g$ is dimensionless constant characterizing the strength of the electron-electron interaction)
 Equations derived in the previous sections are valid (with the replacement of $\gamma$ with $\tilde \gamma$) provided that  inequality \eqref{ineq}
 is satisfied.
  In the opposite limiting case $\tilde \gamma \Delta \ll \Gamma_\varphi^0, $   the    interaction should modify the structure of the resonant peaks  in ${\cal T}(\phi)$ and $P(\phi)$.    One may  expect two effects, predicted previously for the spinless case: \cite{dmitriev} (i) all the resonances  would       acquire fine structure,  splitting into  a series of narrow peaks separated by distance $g$ (ii) each of the peaks of the split structure would broaden within a width $\Gamma_{\varphi}/\Delta,$ where $\Gamma_{\varphi}$ is the phase breaking rate,
   which is expected to be much smaller than the bulk dephasing  rate  ($\Gamma_{\varphi} \ll \Gamma_{\varphi}^0$ )
    because of the charge and size quantization in the almost closed ring.    More detailed analysis including rigorous calculation of
    $\Gamma_{\varphi}$   will be presented elsewhere.\cite{dmitriev1}

    \section{Acknowledgments}
    We thank I.V.~Gornyi  and D.G.~Polyakov for valuable discussions. The work was  supported by Russian Foundation for Basic Research (grants 11-02-00146 and 11-02-91346) and by programs of the Russian Academy od Science.  The work of P.M.~Shmakov was supported by Dynasty Foundation.

\appendix
\section{}
In this appendix we discuss the derivation of the spin rotation matrices \eqref{M}.
 Let us denote as $M(\varphi_0,\varphi)$  the spin rotation matrix, which corresponds  to the counterclockwise trajectory with zero winding number going from $\varphi_0$ to $\varphi.$ To find this matrix we write
 \be
\hat M(\varphi_0,\varphi+\delta\varphi)=\delta \hat M(\varphi)\hat M(\varphi_0,\varphi),\label{Mdelta}
\ee
 where $\delta \hat M(\varphi) =  \hat M(\varphi,\varphi+\delta\varphi)$ describes the spin rotation along  trajectory with an infinitesimally short length.

 In the quasiclassical approximation an electron spin state $\chi$  obeys    the following equation \cite{history1,history6}
 \be
i\hbar\frac{\partial \chi}{\partial \varphi}\frac{v_F}{R} = \hat H^\prime\chi,\label{chi_dif}
\ee
where  $\hat H^\prime$ is the Hamiltonian given by the sum of Eqs.~\eqref{Ham} and \eqref{Rashba}, with the operator $-i\partial/\partial\varphi$  in $\hat H_{SO}$ replaced with $k_F R$ (for $\varphi>\varphi_0$) or with  $-k_F R$ (for $\varphi<\varphi_0$).   From Eq.~\eqref{chi_dif} one can easily conclude that $\delta \hat M(\varphi) = 1-iR\delta\varphi\hat H^\prime/\hbar v_F.$ As a result, we find that the  matrix $\hat M(\varphi_0,\varphi)$ obeys the following differential equation
 \be
\frac{\partial \hat M(\varphi_0,\varphi)}{\partial\varphi} = i\begin{bmatrix} l_z & -l_\perp e^{-i\varphi} \\ -l_\perp e^{i\varphi} & -l_z \end{bmatrix} \hat M(\varphi_0,\varphi) \label{dif_eq}
\ee
with initial condition $\hat M(\varphi_0,\varphi_0)=1.$ Here $l_z = \xi\cos\theta - \Omega_Z$ and $l_\perp=\xi\sin\theta.$ The substitution $$\hat M = \hat M^\prime\begin{bmatrix} e^{-i\varphi/2} & 0 \\ 0 &  e^{i\varphi/2} \end{bmatrix}$$ yields
\be
\frac{\partial \hat M^\prime}{\partial\varphi} = i\begin{bmatrix} l_z+\frac12 & -l_\perp \\ -l_\perp  & -l_z -\frac12\end{bmatrix} \hat M^\prime.\label{dif_eq1}
\ee
Solving \eqref{dif_eq1} we find
\be
\hat M = \begin{bmatrix}
M_{11}
& M_{12}
\\ M_{21}& M_{22}\end{bmatrix},\label{Mphi}
 \ee
 where
 \bee
&&M_{11}=  e^{i{\delta_+(\varphi-\varphi_0)}}\cos^2\frac{\vartheta_+}{2} \nonumber \\
&&~~~~~~~+e^{i{(1+\delta_+ )(\varphi_0-\varphi)}}\sin^2\frac{\vartheta_+}{2}, \nonumber \\
 &&M_{12}=  -i\sin\vartheta_+ \sin[|\varkappa_+|(\varphi-\varphi_0)]e^{-i\frac{\varphi+\varphi_0}{2}},\nonumber  \\
  &&M_{21}=-M_{12}^*,~~~ M_{22}=M_{11}^*.
 %&&M_{21}= -i\sin\vartheta \sin(|\varkappa_+|(\varphi-\varphi_0))e^{i(\varphi+\varphi_0)/2}\nonumber  \\
 %&&M_{22}= e^{-ir_+(\varphi-\varphi_0)/2}\cos^2(\vartheta/2)+e^{i(r_+ +1)(\varphi-\varphi_0)/2}\sin^2(\vartheta/2)
\nonumber
\eee
Here $\delta_+$, $\vartheta_+$ and $\varkappa_+$  are given by Eq.~\eqref{kappa}.

The case $\varphi<\varphi_0$ is considered in an analogous way, leading to the matrix, given by Eq.~\eqref{Mphi}, with the replacements $\delta_+ \to \delta_-, \vartheta_+ \to \vartheta_-$ and $\varkappa_+ \to \varkappa_-.$
%(the difference between two cases is due to  the change of electron velocity direction).
One can check that the matrices $\hat M_{a\to b}^\pm = \hat M(\mp \pi,0)$ are given by Eq.\eqref{M} and that $\hat M_{b\to a}^\pm = \hat M(0,\pm \pi) = (\hat M_{a \to b}^\pm)^T.$

\section{}
In this appendix we derive Eqs.\eqref{T_SOZ}-\eqref{S_SOZ} for the transmission coefficient and spin polarization of the electrons, passing through the ring with the Rashba and Zeeman interactions in the limit of weak coupling to the contacts ($\gamma \ll 1$).

For the purpose of this appendix it is convenient to introduce the probability
  $W_{\chi_i\chi_f} = \big\langle|\langle\chi_f|\hat t|\chi_i\rangle|^2\big\rangle_E$
  %, where $\chi_i$ and $\chi_f$ are arbitrary spinors. This quantity is the probability
%of
for an electron, which approached the contact $a$ in the spin state $|\chi_i\rangle$,  to exit the ring from the contact $b$ in the spin state $|\chi_f\rangle.$

The transmission coefficient and spin polarization are expressed in terms of $W_{\chi_i\chi_f}$ as follows:
\bee
{\cal T} = \frac{W_{\chi_i^\uparrow\chi_f^\uparrow} +W_{\chi_i^\downarrow\chi_f^\uparrow}+W_{\chi_i^\uparrow\chi_f^\downarrow}
+W_{\chi_i^\downarrow\chi_f^\downarrow}}{2},\label{Tfull}\\
\mathbf{Pn} = \frac{W_{\chi_i^\uparrow\chi_{\mathbf n}^\uparrow} +W_{\chi_i^\downarrow\chi_{\mathbf n}^\uparrow}-W_{\chi_i^\uparrow\chi_{\mathbf n}^\downarrow}
-W_{\chi_i^\downarrow\chi_{\mathbf n}^\downarrow}}{2{\cal T}},\label{Pn}
\eee
where $\chi_i^{\uparrow\downarrow}$ and $\chi_f^{\uparrow\downarrow}$ are two arbitrary bases,  $\chi_{\mathbf n}^{\uparrow\downarrow}$ are the eigenstates of the operator $\hat{\boldsymbol\sigma}\mathbf n ~~(\hat{\boldsymbol\sigma}\mathbf n\chi_{\mathbf n}^{\uparrow\downarrow}=\pm\chi_{\mathbf n}^{\uparrow\downarrow})$ and $\mathbf n$ is a unit vector in arbitrary direction.

The probability $W_{\chi_i\chi_f}$ can be found as
\bee
W_{\chi_i\chi_f} &&= \,\sum\limits_{n=0}^{\infty}|\langle\chi_f|\hat\beta_n|\chi_i\rangle|^2   \nonumber\\
&&=  \sum\limits_{n=0}^{\infty}\,\left|\begin{bmatrix} \chi_f \\ \chi_f \end{bmatrix}^\dagger  \hat A^n\begin{bmatrix} \hat \beta_{0}^+ \chi_i\\  \hat \beta_{0}^- \chi_i \end{bmatrix}\right|^2. \label{T_SOZ_sum}
\eee

It can also be written in a form, analogous to Eq.~\eqref{T0_sum2}:
\bee
&&W_{\chi_i\chi_f} = \label{T_kron}\\
&&\left(\begin{bmatrix}\chi_f\\\chi_f \end{bmatrix}\hspace{-1mm}\otimes\hspace{-1mm}\begin{bmatrix}  \hat \beta_0^+\chi_i\\ \hat \beta_0^-\chi_i \end{bmatrix}\right)^\dagger \frac{1}{1-\hat A\otimes \hat A^\dag}\begin{bmatrix} \hat \beta_0^+\chi_i\\ \hat \beta_0^-\chi_i  \end{bmatrix}\hspace{-1mm}\otimes\hspace{-1mm}\begin{bmatrix}   \chi_f\\ \chi_f \end{bmatrix}.\nonumber
\eee
 Though Eq.~\eqref{T_kron} is very useful for numerical analysis it  yields  very cumbersome analytical expressions for ${\cal T}$ and $\mathbf P.$  For the case $\gamma\ll1$ it turns out more convenient to use Eq.~\eqref{T_SOZ_sum} for deriving Eqs.~\eqref{T_SOZ}-\eqref{S_SOZ}.
To this end, we  first   make a  unitary transformation of the matrix $\hat A:$
 \be
\hat A^\prime =  \hat \Lambda \hat A  \hat \Lambda^{-1},
\label{B9}
\ee
described by the block matrix
\be
\hat \Lambda = \begin{bmatrix} \hat \Lambda_+ & 0 \\
0 & \hat \Lambda_-
   \end{bmatrix},\label{B10}
\ee
where $\hat \Lambda_{\pm}$ are $2\times 2$ matrices which correspond  to two unitary transformations diagonalizing matrices $\hat M_+$ and $\hat M_-,$ respectively  (the transformation, described by Eqs.~\eqref{B9}, \eqref{B10}, exactly diagonalizes the matrix $\hat A$ in the case $t_b=0$, when $\hat A$ is a block matrix with blocks $e^{-2 \pi i \phi}\hat M_+$ and $e^{2 \pi i \phi}\hat M_-$).
For weak tunneling coupling, $\gamma \ll 1$, and $t_b \neq 0,$ the matrix $\hat A^\prime$  reads
\begin{widetext}
\be
\hat A^\prime =\frac{1}{(1+\gamma)^2}
\begin{bmatrix}
\lambda_+^2 & 0 & -\gamma c_- \lambda_+(\lambda_+ + \lambda_-) & \gamma s_- \lambda_+ (\lambda_+ - \bar\lambda_-)\\
0 & \bar\lambda_+^2 &  -\gamma s_-\bar\lambda_+(\bar\lambda_+ - \lambda_-)  & -\gamma c_-\bar\lambda_+(\bar\lambda_+ + \bar\lambda_-)\\
-\gamma c_- \lambda_-(\lambda_- + \lambda_+)&-\gamma s_- \lambda_-(\lambda_- -\bar\lambda_+)  &  \lambda_-^2 & 0 \\
\gamma s_- \bar\lambda_-(\bar\lambda_- - \lambda_+) & -\gamma c_- \bar\lambda_-(\bar\lambda_-+\bar\lambda_+) & 0  &  \bar\lambda_-^2
\end{bmatrix},\label{B6}
\ee
where
$\lambda_\pm = e^{\mp i \pi(\phi - \delta_\pm)},$ and $ \bar\lambda_\pm = e^{\mp i \pi(\phi + \delta_\pm)}.$
After the unitary transformation, Eq.~\eqref{T_SOZ_sum} is  rewritten as follows:
\be
 W_{\chi_i\chi_f}
 \approx 4\gamma^2
 \sum\limits_{n=0}^{\infty}\,\left|\begin{bmatrix} \langle\chi_f|\chi_+^1\rangle \\ \langle\chi_f|\chi_+^2\rangle \\ \langle\chi_f|\chi_-^1\rangle \\ \langle\chi_f|\chi_-^2\rangle \end{bmatrix}^\dagger (\hat A^\prime)^n
\begin{bmatrix} \langle \tilde \chi_+^1|\chi_i\rangle \lambda_+\\ \langle \tilde \chi_+^2|\chi_i\rangle \bar\lambda_+ \\ \langle \tilde \chi_-^1|\chi_i\rangle  \lambda_-\\ \langle \tilde \chi_-^2|\chi_i\rangle \bar\lambda_- \end{bmatrix}\right|^2.\label{B5}
\ee
\end{widetext}
 Here
 \be
\chi_\pm^1 = \begin{bmatrix}\cos \vartheta_\pm/2 \\ -\sin \vartheta_\pm/2 \end{bmatrix}, \,\,\, \chi_\pm^2 = \begin{bmatrix}\sin \vartheta_\pm/2 \\ \cos \vartheta_\pm/2 \end{bmatrix}. \label{B1}
\ee
are the eigenstates of the spin rotation matrices $\hat M_+$ and $\hat M_-,$ respectively, and
\be
\tilde\chi_\pm^1 =\begin{bmatrix}\cos \vartheta_\pm/2 \\ \sin \vartheta_\pm/2 \end{bmatrix}, ~~~
\tilde\chi_\pm^2 = \begin{bmatrix}-\sin \vartheta_\pm/2 \\ \cos \vartheta_\pm/2 \end{bmatrix}. \label{B3} %e^{\pm i\pi r_\pm}
\ee
The spinors \eqref{B1} and \eqref{B3} obey ($\mu =1,2$)
 \bee
&&\hat M_\pm \chi_\pm^1 = e^{\pm 2\pi i \delta_\pm} \chi_\pm^1,\nonumber\\
 &&\hat M_\pm \chi_\pm^2 = e^{\mp 2\pi i \delta_\pm} \chi_\pm^2, \label{B2}\\
&& \hat M_{a\to b}^\pm \tilde\chi_\pm^{\mu} =  e^{\mp (-1)^\mu \pi i \delta_\pm}\chi_\pm^{\mu},\nonumber\\
&& \hat M_{b\to a}^\pm \chi_\pm^{\mu} = e^{\mp (-1)^\mu \pi i \delta_\pm}\tilde\chi_\pm^{\mu}. \nonumber
\eee
 Since $\gamma \ll 1,$
 the off-diagonal elements $A_{ij}^\prime$ are small and  should be taken into account  only when they  become comparable with the differences of corresponding diagonal elements: $|A_{ij}^\prime/(A_{ii}^\prime-A_{jj}^\prime)|\gtrsim 1.$
 %(this is analogous to the statement of usual perturbation theory, that the coupling of two levels is proportional to the matrix element of perturbation divided by the energy difference).
 Thus we can neglect the backscattering for all $\phi$, with the exception of the  vicinities of the points, where $\lambda_+ = \lambda_-, \bar\lambda_+ = \bar\lambda_-, \lambda_+ = -\bar\lambda_-$ or $\bar\lambda_+ = -\lambda_-.$ These equations correspond to the magnetic fluxes $ \pm {(\delta_+ + \delta_-)}/{2}$ and $ \pm {(\delta_+ - \delta_-)}/{2}+1/2.$
Hence, the calculations can  be  performed in two steps. First, we solve the problem neglecting the backscattering. The obtained solution is valid everywhere except the vicinities of  points defined above. Next, we assume that the flux is close to one of these points and  take into account backscattering.

The calculation of $W_{\chi_i\chi_f}$ for $t_b=0$ is straightforward, since $A^\prime$ is diagonal, and leads to the results, given by Eqs. \eqref{T_SOZ}-\eqref{S_SOZ} with the substitutions ${\cal T}_0(\phi) \to \mathrm {Re}\, F(\phi)$ and ${\cal T}_0^\prime(\phi) \to \mathrm {Im}\, F(\phi)$, where
\be
F(x) = \frac{8\gamma^2 e^{-2\pi i x}}{1-e^{-4\pi i x}(1+\gamma)^{-4}} +2\gamma\label{B6}
\ee
Both real and imaginary parts of this function have sharp peaks at $x=0$ and at $x=1/2$ with the width on the order of $\gamma$. We see that in this approximation ${\cal T}(\phi)$ and $\mathbf P(\phi)$ have eight resonances per each period,  corresponding to fluxes  $ \pm {(\delta_+ + \delta_-)}/{2}$,  $ \pm {(\delta_+ + \delta_-)}/{2}+1/2$, $ \pm {(\delta_+ - \delta_-)}/{2},$ and $ \pm {(\delta_+ - \delta_-)}/{2}+1/2$. Next we will show that the peaks at $\phi = \pm {(\delta_+ + \delta_-)}/{2}$ and $\phi = \pm {(\delta_+ - \delta_-)}/{2}+1/2$ [which correspond to the peak in $F(x)$ at $x=0$], disappear, when we take the backscattering into account, so only four resonances  in ${\cal T}(\phi)$ and $\mathbf P(\phi)$ remain.
It is worth noting that the latter statement  is true only for the interferometer with equal arms (see Ref.~\onlinecite{dmitriev} for the discussion of the spinless case with unequal arms).

Let us consider, for example, the point $\phi = (\delta_1+\delta_2)/2.$ In the vicinity of this point we can neglect all  off-diagonal elements of the matrix $\hat A^\prime$ except the elements $-\gamma c \lambda_+(\lambda_+ + \lambda_-)$ and $-\gamma c \lambda_-(\lambda_- + \lambda_+)$, so that the matrix $\hat A'$ turns into a block matrix with two $1\times1$ blocks and a $2\times2$ block \cite{comment1}. We can also neglect the interference of the contributions of different blocks when calculating the modulus squared in Eq.~\eqref{B5}, since these interference terms are roughly proportional to $\sum_n (1+\gamma)^{-4n}\exp(i n \Delta\phi)$ with $\Delta\phi\gg \gamma,$\cite{comment1} and therefore are small. The points $\phi = -(\delta_1+\delta_2)/2$ and $\phi = \pm {(\delta_+ - \delta_-)}/{2}+1/2$ are treated in an analogous way.  For each of these points we need to calculate the expression of the following form:
\be
\sum\limits_{n=0}^{\infty}\left|\begin{bmatrix} \zeta_1 \\ \zeta_2  \end{bmatrix}^\dagger
\tilde A^n(\phi^\prime;\eta)
\begin{bmatrix} \zeta_3 e^{-i\pi\phi^\prime}\\ \zeta_4 e^{i\pi\phi^\prime}\end{bmatrix}\right|^2,\label{B7}
\ee
where $\tilde A$ is the $2\times 2$ block of the matrix $\hat A^\prime$:
\bee
&&\tilde A(\phi;\eta)\nonumber\\
&&=\frac1{(1+\gamma)^2}\begin{bmatrix} e^{-2\pi i \phi} & \hspace{-2mm}-\gamma \eta (e^{-2\pi i \phi}+1)\\ -\gamma \eta (e^{2\pi i \phi}+1) & e^{2\pi i \phi}\end{bmatrix}.\nonumber%\label{B7}
\eee
 For example, for the point $\phi = (\delta_1+\delta_2)/2$, the parameters entering Eq.~\eqref{B7}  are as follows: $\phi^\prime =\phi - (\delta_1+\delta_2)/2, \zeta_1 = \langle\chi_f|\chi_+^1\rangle, \zeta_2 =  \langle\chi_f|\chi_-^1\rangle, \zeta_3 = \langle \tilde \chi_+^1|\chi_i\rangle \lambda_+ e^{i\pi\phi^\prime}, \zeta_4 = \langle \tilde \chi_-^1|\chi_i\rangle \lambda_-e^{-i\pi\phi^\prime}, \eta = c_-$ (note that $\zeta_3$ and $\zeta_4$  do not depend on $\phi$).

A convenient way of calculating the expression \eqref{B7} is to rewrite it as an integral
$$%\sum_n |\mathbf v_1, \tilde A^n \mathbf v_2)|^2 =
\int\limits_0^{2\pi} \left|\begin{bmatrix} \zeta_1 \\ \zeta_2  \end{bmatrix}^\dagger
(1- e^{ik}\tilde A)^{-1}
\begin{bmatrix} \zeta_3 e^{-i\pi\phi^\prime}\\ \zeta_4 e^{i\pi\phi^\prime}\end{bmatrix}\right|^2 \frac{dk}{2\pi},\nonumber$$ make a substitution $z=e^{ik}$ and use the identity
\be
(z-\tilde A)^{-1} = \frac{z- \mathrm{Tr} \tilde A +\tilde A}{\mathrm{det}(z-\tilde A)},
\ee
which is valid for any $2\times 2$ matrix.
For  $\phi^\prime \ll 1,$ the calculation   yields
\be
\frac{1}{8\gamma}\left(|k_1|^2 + \frac{\pi^2(\phi^\prime)^2|k_3|^2+ \gamma|\eta k_2 - k_1|^2}{\pi^2(\phi^\prime)^2+\gamma(1-\eta^2)}\right),
\ee
where
\bee
k_1 = \zeta_1^*\zeta_3+\zeta_2^*\zeta_4\nonumber\\
k_2 = \zeta_1^*\zeta_4+\zeta_2^*\zeta_3\nonumber\\
k_3 = \zeta_1^*\zeta_3-\zeta_2^*\zeta_4
\eee

Using this expression, we find $W_{\chi_i\chi_f}$ and get the following simple result for the full transmission coefficient and spin polarization
in the vicinities of $\phi = \pm (\delta_1+\delta_2)/2$ and $\phi = \pm {(\delta_+ - \delta_-)}/{2}+1/2$:
\be
{\cal T} = 2\gamma,\, P_x = P_y = P_z = 0.
\ee
 Thus the backscattering processes destroy the resonances at these points. To write the correct answer, one should replace the function $F(x)$, given by Eq.~\eqref{B6}, with the one that coincides with $F(x)$ for $\gamma \ll x \ll 1-\gamma$, and is constant in vicinity of $x = 0$ and $x=1$. Eqs. \eqref{T00} and \eqref{Tprime} represent, respectively, the real and imaginary parts of such a function.


\begin{thebibliography}{100}

\bibitem{so1} A.~Yacoby,  M.~Heiblum, V.~Umansky, H.~Shtrikman,
D.~Mahalu,   Phys. \ Rev. \ Lett. {\bf73},
3149 (1994).

\bibitem{so2} A.~Yacoby,  M.~Heiblum, D.~Mahalu,  H.~Shtrikman,
  Phys. \ Rev. \ Lett., {\bf 74}, 4047 (1994).

\bibitem{so3} R.~Schuster, E.~Buks, M.~Heiblum, D.~Mahalu, V.~Umansky, H~Shtrikman
 Nature {\bf 385}, 417 (1997).

\bibitem{so5} E.~Buks, R.~Schuster,  M.~Heiblum, D.~Mahalu,  H~Shtrikman,
 Nature {\bf 391}, 871 (1998).

\bibitem{bohm} Y.~Aharonov,  D.~Bohm,  Phys. \ Rev. \ B {\bf 115}, 485 (1959).

\bibitem{aronov} A.G.~Aronov,  Yu.V.~Sharvin,   Rev.\ Mod.\ Phys. {\bf 59}, 755
(1987).

\bibitem{butt} M.~B\"uttiker, Y.~Imry, and M.Ya.~Azbel, Phys.\ Rev.\ A {\bf 30}, 1982 (1984); Y.~Gefen, Y.~Imry, and M.Ya.~Azbel, Phys.\ Rev.\ Lett.\ {\bf 52}, 139 (1984); M.~B\"uttiker,  Y.~Imry, R.~Landauer,  S.~Pinhas, Phys. \ Rev.\ B {\bf 31}, 6207 (1985).

\bibitem{kin} J.M.~Kinaret, M.~Jonson, R.I.~Shekhter, S.~Eggert, Phys. \ Rev.\ B {\bf  57}, 3777 (1998).

\bibitem{jagla} E.A.~Jagla, C.A.~Balseiro, Phys.\ Rev.\ Lett. {\bf 70}, 639
(1993).

\bibitem{dmitriev} A.P.~Dmitriev, I.V.~Gornyi, V.Yu.~Kachorovskii, D.G.~Polyakov
 Phys. \ Rev. \ Lett., {\bf 105}, 036402 (2010).

\bibitem{theorem} Y.~Meir, Y.~Gefen, O.~Entin-Wohlman,    Phys.\ Rev.\ Lett {\bf 63}, 798 (1989).

\bibitem{Stone1} H.~Mathur, A.D.~Stone,  Phys.\ Rev.\ B {\bf 44}, 10957 (1991).

\bibitem{history1} A.G.~Aronov, Y.B.~Lyanda-Geller, Phys.\ Rev.\ Lett {\bf 70}, 343 (1993).

\bibitem{history6} T.Z.~Qian,  Z.B.~Su, Phys.\ Rev.\ Lett.\ {\bf 72}, 2311 (1994).

\bibitem{history2}J.~Nitta, F.E.~Meijer,  H.~Takayanji, Appl.\ Phys.\ Lett {\bf 75}, 695 (1999).

\bibitem{history3} D.~ Frustaglia, K.~ Richter, Phys.\ Rev.\ B {\bf 69}, 235310 (2004).

\bibitem{history4} B.~ Molnar, F.M.~ Peeters, P.~ Vasilopoulos, Phys.\ Rev.\ B {\bf 69}, 155335 (2004).

\bibitem{entin} M.V.~Entin and L.I.~Magarill,  Europhys.\ Lett. {\bf 68} 853 (2004).

\bibitem{Jap} V.~Gritsev, G.~Japaridze, M.~Pletyukhov, D.~Baeriswyl, Phys.\ Rev.\ Lett {\bf 94}, 137207 (2005).

\bibitem{devi} P.~Devillard, A.~Cr\'{e}pieux, K.I.~Imura, and T.~Martin, Phys.\ Rev.\ B {\bf 72}, 041309 (2005).

\bibitem{history5} U.~Aeberland, K.~Wakabayashi, M.~Sigrist, Phys.\ Rev.\ B {\bf 72}, 075328 (2005).

\bibitem{exp1} M.~Konig, A.~Tschetschetkin, E.M.~Hankiewicz, J.~Sinova, V.~Hock, V.~Daumer, M.~Schafer, C.R.~Becker, H.~Buhmann,
 L.W.~Molenkamp, Phys.\ Rev.\ Lett. {\bf 96}, 076804 (2006).

\bibitem{exp2}T.~Bergsten, T.~Kobayashi, Y.~Sekine,  J.~Nitta, Phys.\ Rev.\ Lett. {\bf 97}, 196803 (2006).

\bibitem{citro2} R.~Citro, F.~Romeo, Phys.\ Rev.\ B {\bf 73}, 233304 (2006).

\bibitem{plet} M.~Pletyukhov, V.~Gritsev,  N.~Pauget, Phys.\ Rev.\ B {\bf 74},
045301 (2006).

\bibitem{citro1} R.~Citro, F.~Romeo, Phys.\ Rev.\ B {\bf 74}, 115329 (2006).

\bibitem{kov} A.A.~Kovalev, M.F.~Borunda, T.~Jungwirth, L.W.~Molenkamp, J.~Sinova, Phys. \ Rev. B {\bf 76}, 125307 (2007).

\bibitem{cheng} F.~Cheng, G.~Zhou, J.\ Phys.: Condens.\ Matter {\bf 19}, 136215 (2007).

\bibitem{Romeo} F.~Romeo, R.~Citro,   M.~Marinaro,  Phys.\ Rev.\ B {\bf 78}, 245309 (2008).

\bibitem{lobos} A.M.~Lobos and A.A.~Aligia, Phys.\ Rev.\ Lett {\bf 100}, 016803 (2008).

\bibitem{plet1} M.~Pletyukhov and U.~Z\"ulike, Phys.\  Rev.\  B {\bf 77}, 193304 (2008).

\bibitem{moldov} V.~Moldoveanu and B.~Tanatar, Phys. Rev. B {\bf  81}, 035326 (2010).

\bibitem{Aharony} A.~Aharony, Y.~Tokura, G.Z.~Cohen, O.~Entin-Wohlman,  S.~Katsumoto, Phys.\ Rev.\ B {\bf 84}, 035323 (2011).

\bibitem{dual} C.X.~Liu, J.C.~Budisch, P.~Recher, B.~Trauzettel,  Phys.\ Rev.\ B {\bf 83}, 035407 (2011).

\bibitem{Michetti} P.~Michette and P.~Recher, Phys. Rev. B {\bf 83}, 125420 (2011).

\bibitem{Fermi} We assume here that $E_F$ is the largest energy scale in the problem, thus replacing  in the expression for $\Omega_Z$ the  energy-dependent electron velocity $v$ with $v_F.$

\bibitem{AC} Y.~Aharonov, A.~Casher, Phys.\ Rev.\ Lett {\bf 53}, 319 (1984)

\bibitem{Stone} H.~Mathur, A.D.~Stone, Phys.\ Rev.\ Lett {\bf 68}, 2964 (1992)


\bibitem{berry} M.V.~Berry, Proc. \ R. \ Soc.\ London \ A {\bf 392}, 45 (1984).

\bibitem{comment2} Beyond the adiabatic limit, the  AC phase can be also separated  into the dynamical part and  the geometrical Aharonov-Anandan phase. \cite{AA} The latter is a generalization of the Berry geometrical phase for the nonadiabatic case (see discussion in Ref.~\onlinecite{history6})


\bibitem{AA} Y.~Aharonov and J.~Anandan, Phys. \ Rev. \ Lett. {\bf 58}, 1593 (1987).


\bibitem{rao1} S.~Lal, S.~Rao, and D.~Sen, Phys. \ Rev. \ B {\bf 66}, 165327
(2002).

\bibitem{rao2} S.~Das, S.~Rao, and D.~Sen, Phys. \ Rev. \ B {\bf 70}, 085318
(2004).

\bibitem{aristov} D.N.~Aristov, A.P.~Dmitriev, I.V.~Gornyi, V.Yu.~Kachorovskii, D.G.~Polyakov, and P.~W\"{o}lfle, Phys.\ Rev.\ Lett., {\bf 105}, 266404 (2010).
\bibitem{aristov1} D.N.~Aristov,   Phys.\ Rev.\ B  {\bf 83}, 115446 (2011).
\bibitem{aristov2} D.N.~Aristov  and P.~W\"{o}lfle,  Phys.\ Rev.\ B {\bf 84}, 155426 (2011).


\bibitem{dmitriev1} A.P.~Dmitriev, I.V.~Gornyi, V.Yu.~Kachorovskii, D.G.~Polyakov, to be published.


\bibitem{yashenkin} A.G.~Yashenkin, I.V.~Gornyi, A.D.~Mirlin, and D.G.~Polyakov, Phys. \ Rev. \ B {\bf 78}, 205407 (2008).


\bibitem{comment1} We restict our analysis to the case, when the four points $\phi = \pm {(r_+ + r_-)}/{2}$ and $\phi = \pm {(r_+ - r_-)}/{2}+1/2$ are sufficiently far away from each other.


\end{thebibliography}
\end{document}